\def\be{\begin{equation}}
\def\eq{\end{equation}}
\newcommand{\bea}{\begin{eqnarray}}
\newcommand{\eea}{\end{eqnarray}}
\newcommand{\ba}{\begin{eqnarray*}}
\newcommand{\ea}{\end{eqnarray*}}
\newcommand{\nn}{\nonumber}

\documentclass[aps,prd,showpacs,twocolumn,preprintnumbers]{revtex4}

\usepackage{amsmath}
\usepackage{amsfonts} 
\usepackage{amssymb}
\usepackage{xcolor}
\usepackage{graphicx}
\usepackage{dcolumn}
\usepackage{bm}
\usepackage{verbatim}
\usepackage{float}
\usepackage{slashed}
\usepackage{appendix}

\begin{document}

\preprint{EFI 08-11}
\preprint{MCTP 08-14}

\title{Interaction of Dirac and Majorana Neutrinos with Weak Gravitational Fields.}
\author{A. Menon}
\email{aamenon@umich.edu}
\affiliation{Michigan Center for Theoretical Physics and Department of Physics, University of Michigan, 500 E. University Ave., MI 48109-1120}
\author{Arun M. Thalapillil}
\email{madhav@uchicago.edu}
\affiliation{Enrico Fermi Institute and Department of Physics, University of Chicago, 5640 South Ellis Avenue, Chicago, IL 60637\\}

\date{\today}

\begin{abstract}
In this paper the interaction of high energy neutrinos with weak gravitational fields is briefly explored. The form of the graviton-neutrino vertex is motivated from Lorentz and gauge invariance and the non-relativistic interpretations of the neutrino gravitational  form factors are obtained. We comment on the renormalization conditions, the preservation of the weak equivalence principle and the definition of the neutrino mass radius. We associate the neutrino gravitational form factors with specific angular momentum states. Based on Feynman diagrams, spin-statistics, CP invariance and symmetries of the angular momentum states in the neutrino-graviton vertex, we deduce differences between the Majorana and Dirac cases. It is then proved that in spite of the theoretical differences between the two cases, as far as experiments are considered, they would be virtually indistinguishable for any space-time geometry satisfying the weak field condition. We then calculate the transition gravitational form factors for the neutrino by evaluating the relevant Feynman diagrams at 1-loop and estimate a neutrino transition mass radius. The form factor is seen to depend on the momentum transfer very weakly. It is also seen that the neutrino transition mass radius is smaller than the typical neutrino charge radius by a couple of orders of magnitude.
\end{abstract}
\pacs{13.15.+g, 04.25.Nx, 14.60.Lm}

\maketitle

\begin{section}{Introduction}
Neutrinos are very weakly interacting particles, experiencing only the weak nuclear force and the gravitational force. They are produced copiously in many high energy astrophysical processes and due to their weakly interacting nature travel almost unhindered to earth. These characteristics make them a potentially powerful probe of astrophysical phenomena which are otherwise inaccessible. It is speculated that these ultra high energy neutrinos may have energies as high as $\sim10^{12}\text{GeV}$. (See for example~\cite{astroneu} and references therein.)  Many ongoing and planned experiments aim to detect these high energy astrophysical neutrinos~\cite{anexp}. Results from these neutrino telescopes have the potential to provide deep insights in cosmology and particle physics.
\par
Our aim is to study how high energy neutrinos, with $E_{\nu}\gg m_{\nu}$, interact with gravitational fields that are weak. Many of the gravitational systems that we are generally interested in have weak gravity. To our knowledge the first study on neutrino gravity interactions was that of Brill and Wheeler~\cite{brwh} in the context of introducing spinors in general relativity. There have been subsequent studies on neutrino-gravitational effects, mainly pertaining to neutrino oscillations.
\par
 One approach pioneered in~\cite{voep} was to assume that the weak equivalence principle is slightly violated in nature, meaning that the gravitational coupling constant is non-universal and depends on the flavor of the particle species, and then probe its effect on oscillations. Another approach has been to define a covariant gravitational phase~\cite{gphase} (in the context of the Dirac equation in curved space-time) that would be the relevant substitute for the vacuum phase $e^{ipx}$ in curved space-times and explore its effects on neutrino oscillations. There have also been detailed studies of gravitational effects on neutrino oscillations in a medium~\cite{pbpjn} and on spin oscillations in an external gravitational field~\cite{maximdv}.
\par
Towards the goal of understanding neutrino gravity interactions we first study the graviton-neutrino vertex using general symmetry principles and then explore the non-relativistic limits of the neutrino-graviton form factors. This line of reasoning was motivated by the use of invariance principles, long back, in studying the photon-neutrino coupling~\cite{kay}. We use arguments from Feynman diagrams, Fermi-Dirac statistics, CP invariance, crossing symmetry and symmetries of the angular momentum states to study the vertex. This would, among other things, give us some insight into how the Majorana neutrino and Dirac neutrino cases could be different as far as gravitational interactions are concerned. Remarks are also made on the weak equivalence principle and its validity in the context of renormalization. It must be pointed out that some of the differences between Majorana and Dirac form factors were anticipated earlier in~\cite{klak} using methods different from the one we follow. 
\par
In spite of theoretical differences, we prove to $\mathcal{O}(m_{\nu}/E_{\nu})$ that the Majorana and Dirac cases cannot be distinguished by gravitational interactions, as far as experiments are involved. This result is found to hold true for any space-time geometry that satisfies the weak field criterion and is a simple extension of the \textit{practical Majorana-Dirac Confusion theorems}~\cite{kay} first discussed in the context of neutrino scattering and neutrino-photon interactions. 
\par
We also perform an approximate calculation of the off-diagonal (in mass basis) neutrino-graviton form factors at 1-loop to understand their $q^{2}$ dependence. Using the loop calculation results the $\nu_{i}\rightarrow\nu_{j}$ neutrino transition mass radius is estimated when $i\neq j$. A comparison is then made of the neutrino transition mass radius and the neutrino transition charge radius. Finally, the difference in order of magnitudes of the two radii is motivated based on some physical arguments.
\end{section}
\begin{section} {The Neutrino-Graviton Vertex}
For weak gravitational fields we may expand the metric as
\be
\label{linmet}
g_{\mu\nu}\simeq\eta_{\mu\nu}+\kappa h_{\mu\nu}+\mathcal{O}(h^{2})
\eq
where  $\kappa=\sqrt{32\pi G}$, $\eta_{\mu\nu}=(1,-1,-1,-1)$ and $h^{\mu\nu}$ is interpreted as the spin-2 graviton. $G$ is the Newton's gravitational constant. The usual criterion for considering a gravitational field to be weak is $\delta_{g} \sim |\kappa h^{\mu\nu}|\ll1$~\cite{grb}. For the case of spherical symmetry, the exterior is the Schwarzschild spacetime. In this particular case
\ba
\delta_{g} \simeq\frac{2GM}{c^{2} R}
\ea
where $M$ is the mass of the body, $R$ is the radial distance and $c$ is the speed of light.
Let us consider some typical examples now. For the exterior of the sun the above expression gives
\ba
\delta_{\odot}\approx~10^{-6}~~~.
\ea
Active galactic nuclei typically have $M\sim10^{12} M_{\odot}$ and $R\sim10\text{ KPc}$ which means
\ba
\delta_{AGN}\approx~10^{-5}~~~.
\ea
A typical neutron star mass and radius, range between $1.3-2.1~M_{\odot}$  and $10-20~\text{Km}$\,\cite{grb}. This typically leads to
\ba
\delta_{N}\approx~0.3~~~.
\ea
Thus we may be optimistic that the analysis we perform is valid to a good extent in many astrophysical cases of interest.  
\par
The graviton-neutrino vertex current $\mathcal{J}_{\mu\nu}$ in Fig. (\ref{gncpl}) transforms as a symmetric rank-2 tensor. The curly line denotes the off-shell graviton. The spin-2 gauge current is just the energy-momentum tensor of the neutrino in the gravitational background. By Lorentz invariance the possible vertex factors are
\bea 
\nn
&&\bar{u}( p^{'})_{i}i\mathcal{J}_{\mu\nu}u(p)_{i}=\bar{u}( p^{'})_{i}\Big[\hat{A}(q^{2})g_{\mu\nu}+\hat{B}(q^{2})(r_{\mu}r_{\nu})~~~\\ \nn
&&+\hat{C}(q^{2})(q_{\mu}q_{\nu})+ \hat{D}(q^{2})(\gamma_{\{\mu}r_{\nu\}})+\hat{E}(q^{2})(\gamma_{\{\mu}q_{\nu\}})\\ 
&&+\hat{F}(q^{2})(q_{\{\mu}r_{\nu\}})\Big]u(p)_{i}
 \label{gff}
\eea
where $p$ is the incoming neutrino momenta, $ p^{'}$ is the outgoing neutrino momenta, $q= p^{'}-p$, $r= p^{'}+p$ and $\{ \}$ denotes complete symmetrization of the indices. The subscript $i$ in the spinors label the mass eigenstate of the incoming and outgoing neutrinos.
\par
 Using the Slavnov-Taylor-Ward identities 
 \ba
 q^{\mu}~\bar{u}( p^{'})_{i}~i\mathcal{J}_{\mu\nu}~u(p)_{i}=~0
 \ea
 to impose gauge invariance in Eq. (\ref{gff}) we find that the vertex in general must have the form
  \bea
\nn
&&\bar{u}( p^{'})_{i}i\mathcal{J}^{'}_{\mu\nu}\left[i,i\right] u(p)_{i}=\bar{u}( p^{'})_{i}\Big[F_{1}(q^{2})(q^{2} g_{\mu\nu}-q_{\mu}q_{\nu})~~~~\\ 
&&+ F_{2}(q^{2})(r_{\mu}r_{\nu})+ F_{3}(q^{2})(\gamma_{\{\mu}r_{\nu\}})\Big] u(p)_{i}~~~.
 \label{Fforms}
\eea
\begin{figure}
\begin{center}
\includegraphics[width=5.5cm,angle=0]{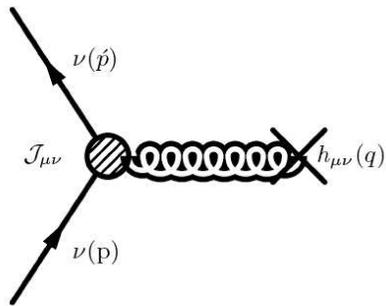}
\end{center}
\caption{Neutrino Graviton Vertex.}
\label{gncpl}
\end{figure}
To arrive at the above expression we used the Dirac equation along with the result $r\cdot q=0$. The notation $\left[i,i\right] $ denotes the mass diagonal case. Also, we will label the parity conserving and parity violating currents by $\mathcal{J}^{'}_{\mu\nu}$ and $\mathcal{J}^{''}_{\mu\nu}$ respectively. The form factors we have defined above may be shown to be linear combinations of the ones defined, for the mass-diagonal case,  in \cite{Nieves:2007jz}  in the context of studying the gravitational decay of a spin-1/2 particle. 
\par
It will be seen that when the relevant graviton Feynman rules are derived to lowest order in $\kappa$, only the $\gamma_{\{\alpha}r_{\beta\}}$ operator appears at tree level. Also note that the operator $r_{\mu}r_{\nu}$ is a chirality flipping operator. Due to the tiny neutrino masses and the left handed nature of the electroweak currents, this would imply that this operator is suppressed by $m_{\nu}/E_{\nu}$. 
\par
Until this point we were considering only parity conserving operators. If we include operators with $\gamma_{5}$ and proceed as above, we find that there are again three possible vertex operators
\bea
\nn
&&\bar{u}( p^{'})_{i}i\mathcal{J}^{''}_{\mu\nu}\left[i,i\right]u(p)_{i}=\bar{u}( p^{'})_{i}\Big[G_{1}(q^{2})(q^{2} g_{\mu\nu}-q_{\mu}q_{\nu})\gamma^{5}\\ \nn
&&+G_{2}(q^{2})(r_{\mu}r_{\nu})\gamma^{5}+G_{3}(q^{2})\lbrace q^{2}(\gamma_{\{\mu}r_{\nu\}})\\
&&- 2m_{\nu_{i}}(q_{\{\mu}r_{\nu\}})\rbrace\gamma^{5}\Big]~u(p)_{i}~~~. \label{Gforms}
\eea

\par
For the off-diagonal (in mass eigenstates) case $\nu_{i}\rightarrow\nu_{j}$ with mass eigenstate $i\neq j$ similar arguments immediately give
\bea
\nn
&&\bar{u}( p^{'})_{j}i\mathcal{J}^{'}_{\mu\nu}\left[i,j\right]u(p)_{i}\simeq\bar{u}( p^{'})_{j}\Big[E_{1}(q^{2})(q^{2} g_{\mu\nu}-q_{\mu}q_{\nu})~~~~~~\\ \nn
&&+E_{2}(q^{2})(r_{\mu}r_{\nu})+E_{3}(q^{2})\lbrace q^{2}(\gamma_{\{\mu}r_{\nu\}})-\Delta_{ij}m_{\nu}(q_{\{\mu}r_{\nu\}})\rbrace\Big]\\ 
&& u(p)_{i}+\mathcal{O}(\Delta_{ij}m^{2}_{\nu})~~~,
\label{Eforms}
\eea

\bea
\nn
&&\bar{u}( p^{'})_{j}i\mathcal{J}^{''}_{\mu\nu}\left[i, j\right]u(p)_{i}\simeq\bar{u}( p^{'})_{j}\Big[ D_{1}(q^{2})\gamma^{5}\\ \nn
&&(q^{2} g_{\mu\nu}-q_{\mu}q_{\nu})+D_{2}(q^{2})(r_{\mu}r_{\nu})\gamma^{5}+D_{3}(q^{2})\lbrace q^{2}(\gamma_{\{\mu}r_{\nu\}})\\ 
&-&\Sigma_{ij}m_{\nu}(q_{\{\mu}r_{\nu\}})\rbrace\gamma^{5}\Big] u(p)_{i}+\mathcal{O}(\Delta_{ij}m^{2}_{\nu})~~~,
\label{Dforms}
\eea
where the notation $\left[i,j\right] $ denotes mass off-diagonal transitions. Also, we define $\Delta_{ij}m_{\nu}=m_{\nu_{j}}-m_{\nu_{i}} $, $\Sigma_{ij}m_{\nu}=m_{\nu_{j}}+m_{\nu_{i}} $ and $\Delta_{ij}m^{2}_{\nu}=m^{2}_{\nu_{j}}-m^{2}_{\nu_{i}} $. In writing the above expressions for the off-diagonal case we have neglected terms of $\mathcal{O}(\Delta_{ij}m^{2}_{\nu})$. Since we are  not interested in any neutrino oscillation phenomena this is justified. For the exact vertex in the case of $i\neq j$ the reader is referred to \cite{Nieves:2007jz}.
\par
Let us now try to deduce physical interpretations for the neutrino form factors in Eqs.\,(\ref{Fforms}) and (\ref{Gforms}) in the non-relativistic limit. In the electromagnetic case it is well known that in the non-relativistic limit the form factors are interpreted as charge and electromagnetic moments. For a neutrino in that case the neutrino charge vanishes. It is thus interesting to see what the interpretations would be for the neutrino-graviton case. It must be mentioned that the energy-momentum structure form factors and their non-relativistic interpretations were first studied in the context of understanding the `mechanical structure' of particles and the self-energy of the electron~\cite{hp}. 
\par
Consider the neutrino form factors in the limit $q\rightarrow 0$. Then we have the graviton-neutrino interaction Hamiltonian density
\ba
&&\lim_{q\rightarrow0}\mathcal{H}_{g}=\kappa h^{\mu\nu}~\bar{u}(p)_{i} i\mathcal{J}_{\mu\nu} u(p)_{i}\\ \nn
&&\simeq\kappa h^{\mu\nu}\bar{u}(p)_{i} [ 4(F_{2}(0)+G_{2}(0)\gamma^{5})p_{\mu}p_{\nu}+2 F_{3}(0)\gamma_{\{\mu}p_{\nu\}}] \\ \nn
&&u(p)_{i}~~~.
\ea
Further simplification is achieved if we transform into the rest frame of the incoming neutrino. In the rest frame the only non-zero component of $\mathcal{J}_{\mu\nu}$ is $\mathcal{J}_{00}$ and the above expression reduces to
\ba
&&\kappa~h^{00} \bar{u}(\vec{p}\rightarrow0)_{i}~i\mathcal{J}_{00}~u(\vec{p}\rightarrow0)_{i}\simeq\kappa~h^{00} \bar{u}(\vec{p}\rightarrow0)_{i}\\ \nn
&&\Big[4(F_{2}(0)+G_{2}(0)\gamma^{5}) m^{2}_{\nu_{i}}+4 F_{3}(0)m_{\nu_{i}}\Big]~u(\vec{p}\rightarrow0)_{i}
\ea
Based on the above expressions let us re-define the form factors with appropriate normalization factors so as to obtain a simpler result,
\ba
F_{2}(q^{2})\rightarrow \frac{F_{2}(q^{2})}{4 m_{\nu_{i}}},~G_{2}(q^{2})\rightarrow \frac{G_{2}(q^{2})}{4 m_{\nu_{i}}},~F_{3}(q^{2})\rightarrow \frac{F_{3}(q^{2})}{4}~.
\ea
In the above expressions, it is assumed that $m_{\nu_{i}}$ are neutrino masses measured from some kinematics. Thus they may be regarded as neutrino masses in the `inertial' sense.
With this new normalization we then have
\bea
\label{zcpl}
&&\kappa h^{00}\bar{u}(0)_{i}i\mathcal{J}_{00}u(0)_{i}\approx m_{\nu_{i}}\Phi_{\text{\tiny{g}}}\left[F_{2}(0)+F_{3}(0)\right]\phi^{\dag}\phi~~~~
\eea
where $\phi$ is the usual ``large component" Pauli spinor (of the neutrino) and $\Phi_{\text{\tiny{g}}}$ is the gravitational potential in the neutrino rest frame. To arrive at the above expression we used the well known fact that in the weak gravity limit
\ba
g^{00}\simeq1+2\Phi_{\text{\tiny{g}}}
\ea
 in natural units. 
 \par
 From Eq. (\ref{zcpl}) we deduce that $F_{2}(q^{2})$ and $F_{3}(q^{2})$ are associated with the energy density or mass of the incoming neutrino. This observation leads to the requirement that in the limit $q\rightarrow0$ we must have 
\be
\label{wep}
F_{2}(0)+~F_{3}(0)=~1~~~.
\eq
\par
 But we already noted that $r_{\mu}r_{\nu}$ is a neutrino chirality flipping operator. 
 In the $q\rightarrow 0$ limit $F_{2}(q^{2})$ must vanish, since in the limit of zero momentum transfer we do not want to induce any neutrino chirality flips. This may be considered as a consequence of angular momentum conservation. Hence,
\be
 F_{2}(0)\rightarrow0~~~.
 \eq
 \par
 The above conditions imply that any correction, to the neutrino mass, calculated at one-loop be renormalized such that in the limit  $q\rightarrow0$ it vanishes:
 \bea
 \nn
\Delta\hat{F}_{i}(q^{2})_{\text{\tiny{1-loop}}}&=& \Delta F_{i}(q^{2})_{\text{\tiny{1-loop}}}-\Delta F_{i}(0)_{\text{\tiny{1-loop}}}\\
\Delta\hat{G}_{i}(q^{2})_{\text{\tiny{1-loop}}}&=& \Delta G_{i}(q^{2})_{\text{\tiny{1-loop}}}-\Delta G_{i}(0)_{\text{\tiny{1-loop}}}~.
\label{mreno}
 \eea
 This renormalization condition is exactly analogous to the electromagnetic case where in the limit $q\rightarrow0$ we require the one-loop vertex corrections to not renormalize the electric charge or moments.  
  \par
Note also that for the off-diagonal case when $i\neq j$ such a renormalization must again be imposed as $q\rightarrow 0$ to prevent the generation of off-diagonal  elements in the non-relativistic limit. This condition will be used explicitly later in section IV when we calculate the mass off-diagonal form factors.
\par
A version of the \textit{weak equivalence principle} states that the gravitational mass is equal to the inertial mass for all particles~\cite{grb}. We see that, imposing the conditions in Eq.\, (\ref{mreno}) at every step ensures that the mass eigenstates satisfy the weak equivalence principle. This would also mean, among other things, that a weak gravitational field does not affect neutrino oscillations. The relevant oscillation in pure gravity (weak) would still be the vacuum oscillation. 
 \par
 Analogous to the electromagnetic case, one can now consider the $F_{3}(q^{2})$ form factor roughly as the Fourier transform of a hypothetical neutrino mass density $\rho^{\text{\tiny{$\nu$}}}_{\text{\tiny{m}}}(r)$, 
 \ba
 F_{3}(q^{2}) \sim\int \rho^{\text{\tiny{$\nu$}}}_{\text{\tiny{m}}}(r) e^{i\vec{q}\cdot\vec{r}} d^{3}r~~~.
 \ea
 This leads naturally to a definition of the \textit{neutrino mass radius} 
 \be
 \big\langle r^{2}_{\nu}\big\rangle_{m}\sim -6\frac{d}{dq^{2}}\left[F_{3}(q^{2})\right]\bigg\vert_{q^{2}\rightarrow 0}~~~.
 \label{crdef}
 \eq
 Later, we will estimate the $i\rightarrow j$ neutrino transition mass radius by explicitly evaluating the Feynman diagrams at 1-loop.
 \par
 Let us now generalize the previous results for the neutrino form factors. By the virial theorem, for any energy-momentum tensor we have the general result~(see for example \cite{grb})
 \ba
 \int~d^{3}x~\langle\mathcal{J}^{\alpha}_{\alpha}\rangle_{\text{t}}= \int~d^{3}x~\langle\mathcal{J}^{0}_{0}\rangle_{\text{t}}=\langle E\rangle_{\text{t}}
 \ea
 where $\langle E\rangle_{\text{t}}$ is the time-averaged energy and $\langle\rangle_{\text{t}}$ denotes time averaging.
This implies that
\ba
\langle\mathcal{J}^{\alpha}_{\alpha}\rangle_{\text{t}}=\langle\rho_{E}\rangle_{\text{t}}
\ea
is the average energy density. So if we are considering neutrino interactions in an environment of gravitons, in the non-relativistic limit, we may use the above expression in Eqs.\,(\ref{Fforms}) and (\ref{Gforms}) along with the normalizations and interpret
\bea
\nn
&&\langle\rho^{\nu,h}_{E}\rangle_{\text{t}}\approx\bigg<m_{\nu}\left[F_{2}(q^{2})+F_{3}(q^{2})\right]\phi^{\dag}\phi\bigg >_{\text{t}}\\ 
&&+\bigg<q^{2}\left[4F_{1}(q^{2})-\frac{F_{2}(q^{2})}{4m_{\nu}}\right]\phi^{\dag}\phi \bigg >_{\text{t}}
\label{aved}
\eea
as an \textit{average energy density for the incoming neutrino}. The first term is familiar from the $q\rightarrow0$ case and the second term seems to give a next order correction in $q^{2}$.
 \par
All the above expressions give an interpretation for the neutrino form factors $F_{1}(q^{2}),\,F_{2}(q^{2})$ and $F_{3}(q^{2})$. The form factors $G_{1}(q^{2}),\,G_{2}(q^{2})$ and $G_{3}(q^{2})$ may be thought of as the `axial' counterparts of the $F(q^{2})$ form factors. To summarize, the complete mass-diagonal graviton-neutrino vertex is of the form
\bea
\nn
&&\bar{u}( p^{'})_{i}i\mathcal{J}_{\mu\nu}u(p)_{i}=\bar{u}( p^{'})_{i}\Big[F_{1}(q^{2})(q^{2} g_{\mu\nu}-q_{\mu}q_{\nu})~~~\\ \nn
&&+\frac{F_{2}(q^{2})}{4m_{\nu}}(r_{\mu}r_{\nu})+\frac{F_{3}(q^{2})}{4}(\gamma_{\{\mu}r_{\nu\}})\\ \nn
&&+G_{1}(q^{2})(q^{2} g_{\mu\nu}-q_{\mu}q_{\nu})\gamma^{5}+\frac{G_{2}(q^{2})}{4m_{\nu}}(r_{\mu}r_{\nu})\gamma^{5}\\ 
&&+ G_{3}(q^{2})\lbrace q^{2}(\gamma_{\{\mu}r_{\nu\}})- 2m_{\nu_{i}}(q_{\{\mu}r_{\nu\}})\rbrace\gamma^{5}\Big]   u(p)_{i}\, ,
\label{fullforms}
\eea
with the interpretations given above.
\par
Further relations between the form factors may now be obtained by including information about the nature of the standard model gauge interactions. To be specific, if we incorporate the observational fact that electroweak currents are left-handed and neutrino masses are very tiny then by considering the matrix elements between various chiral states we can further constrain the form factors. We will see later on that at tree level the only operator that appears is $\gamma_{\{\alpha}r_{\beta\}}$. The other factors are generated at higher orders, so we have the obvious constraints on them from electroweak theory that matrix elements of right-right chirality or right-left chirality vanish owing to the tiny neutrino masses. For example
\bea
\nn
\langle\nu_{R}\lvert \Delta\mathcal{J}_{\mu\nu}\rvert\nu_{R}\rangle \xrightarrow{m_{\nu}\rightarrow 0 }~0 \\ \nn
\eea
would imply (choosing an appropriate non-chiral Dirac basis) that
\ba
F_{i\neq3}(q^{2})&\simeq& G_{i\neq3}(q^{2}) \\ \nn
\Delta F_{3}(q^{2})&\simeq& \Delta G_{3}(q^{2}) \nn
\ea
to $\mathcal{O}(m_{\nu}/E_{\nu})$.
\par
A similar line of reasoning will be used when we address what the experimental consequences of the results we obtain are, for distinguishing Majorana and Dirac neutrinos through gravitational interactions.
\par
Later in our analysis we will also explicitly take the non-relativistic limit of the operators in Eq.\,(\ref{fullforms})  and identify the angular momentum states associated with each one of them. This along with other arguments from spin-statistics, CP, crossing symmetry and Feynman diagrams will yield information about how the Majorana and Dirac cases differ as far as neutrino gravitational interactions are concerned. 
\section{Angular Momentum states and Symmetries of the vertex }
\par
  Let us now consider the hypothetical case of identical Majorana fermions in the final state, for the s-channel process with matrix elements $\langle\nu( p^{'})_{i}\nu(p)_{i}\lvert\mathcal{J}_{\mu\nu}\rvert0\rangle$, which is related to the t-channel process (through crossing symmetry) with matrix elements $\langle\nu( p^{'})_{i}\lvert\mathcal{J}_{\mu\nu}\rvert\nu(p)_{i}\rangle$. Explicitly the process under consideration is 
\ba
h^{*}(q)~\rightarrow~\nu_{i}(p)~+~\nu_{i}( p^{'})~~~.
\ea
In the above, $h^{*}$ is the \textit{off-shell} graviton associated with the weak gravitational field. For the s-channel process we have the new definitions $q= p^{'}+p$ and $r= p^{'}-p$. As an aside, note that unlike processes such as $\nu\rightarrow \nu+\gamma$ where the photon may be on-shell, in a t-channel process $\nu_{i}+h\rightarrow \nu_{j}$ the spin-2 graviton cannot in general be on-shell, due to angular momentum conservation. The possibility of an on-shell graviton considered by K.L. Ng in \cite{klak} seems to be in disagreement with this reasoning.

\par
For exploring the differences between graviton-Majorana neutrino and graviton-Dirac neutrino form factors, we will adopt a line of reasoning similar to the one used earlier in the context of neutrino photon couplings. In the case of photon couplings, Fermi-Dirac statistics and some general results were used to show that a Majorana neutrino can have only a single electromagnetic form factor compared to four in the Dirac case\,\cite{kay}.
\par
Henceforth, a super-script $D$ will label the Dirac neutrino case and $M$ will label the Majorana neutrino case. 
 \par
Assume that the final state fermions in our case are identical Majorana neutrinos. By Fermi-Dirac statistics we must then have 
\bea
\nn
&&\langle\nu( p^{'})_{i}\nu(p)_{i}\lvert\mathcal{J}_{\mu\nu}\rvert0\rangle=\bar{u}_{i}( p^{'})i\mathcal{J}^{M}_{\mu\nu}\left[(p^{'}, s^{'}),(p,s)\right]v_{i}(p)\\ 
&&=-\bar{u}_{i}(p)i\mathcal{J}^{M}_{\mu\nu}\left[(p,s), ( p^{'}, s^{'})\right]v_{i}( p^{'})~~~.
\label{fdmaj}
\eea
We note that under this exchange $q\rightarrow q$ while $r \rightarrow -r$ in $\mathcal{J}_{\mu\nu}$. In addition to the above equation we also have
\bea
\nn
&&\bar{u}( p^{'}, s^{'})i\mathcal{J}^{\mu\nu}v(p,s)=[i\gamma_{2}v( p^{'})]^{T}\gamma_{0}i\mathcal{J}^{\mu\nu}[-i\gamma_{2}u^{*}(p)]~~~~\\ 
&&\equiv \bar{u}(p)\left[\hat{\mathcal{C}}i\mathcal{J}^{T}_{\mu\nu}\hat{\mathcal{C}}\right]v( p^{'})=-\bar{u}(p)\left[\hat{\mathcal{C}}^{-1}i\mathcal{J}_{\mu\nu}\hat{\mathcal{C}}\right]^{T}v( p^{'})
\label{cmaj}
\eea
where $\hat{\mathcal{C}} =\gamma^{0} \gamma^{2} $. For Majorana neutrinos, from Eqs. (\ref{fdmaj}) and (\ref{cmaj}), the s-channel matrix elements must satisfy
\ba
&&\langle\nu( p^{'})_{i}\nu(p)_{i}\lvert\mathcal{J}_{\mu\nu}\rvert0\rangle=\bar{u}_{i}( p^{'})i\mathcal{J}^{M}_{\mu\nu}\left[( p^{'}, s^{'}),(p,s)\right]v_{i}(p)~~~~~\\ \nn
&&=-\bar{u}(p)_{i}i\mathcal{J}^{M}_{\mu\nu} \left[(p,s),( p^{'}, s^{'})\right]v( p^{'})_{i}\\ \nn
&&=-\bar{u}(p)_{i}\left[\hat{\mathcal{C}}^{-1}i\mathcal{J}^{M}_{\mu\nu}\left[( p^{'}, s^{'}), \left(p,s\right)\right]\hat{\mathcal{C}}\right]^{T} v( p^{'})_{i}\nn~~~.
\ea
Thus we may conclude from above that for a Majorana neutrino mass eigenstate the only form factors that survive are those for which
\be
\label{majcond}
\left[\hat{\mathcal{C}}^{-1}\mathcal{J}^{M}_{\mu\nu}\left[( p^{'}, s^{'}),(p,s)\right]\hat{\mathcal{C}}\right]^{T}=\mathcal{J}^{M}_{\mu\nu}\left[\left(p,s\right),( p^{'}, s^{'})\right]~~~.
\eq
It is to be emphasized that the above conclusion is derived solely from Fermi-Dirac statistics and the fact that for Majorana neutrinos the final state in the equivalent s-channel process consists of identical fermions labeled by $i$. Using the above criterion on the terms in Eq. (\ref{fullforms}) it is found that for the Majorana case the form factor

\be
G^{M}_{3}(q^{2})~=~0~~~.
\label{gm3}
\eq
This was also noted previously, following a different method, in ~\cite{klak}. Thus we may claim that, \emph{in the coupling of a neutrino to a graviton, the Dirac neutrino has \textit{six} non vanishing form factors while the Majorana neutrino has only \textit{five} form factors to be consistent with Fermi-Dirac statistics.}
\par
We also note that for the cases when there are transitions $\nu^{M}_{i}\rightarrow\nu^{M}_{j}$ with $i\neq j$, by crossing symmetry, we have a correspondence between the processes
\ba
\nu^{M}_{i}\xrightarrow{h^{*}} \nu^{M}_{j}~\Leftrightarrow~h^{*}\longrightarrow \nu^{M}_{i}+\nu^{M}_{j}~ \Leftrightarrow ~\nu^{M}_{j}\xrightarrow{h^{*}} \nu^{M}_{i}~,
\ea
implying that the  \textit{form factors of these matrix elements are the same}.
\par
A more transparent way to understand why the form factor $G^{M}_{3}(q^{2})$ vanishes in the Majorana case is by looking at the angular momentum states (of the final neutrinos) produced by the virtual massless graviton. This method also leads to other insights regarding neutrino gravitational interactions. 
\par
A virtual massless graviton has a $J=0$ as well as a $J=2$ component~\cite{ddsw}. This is very different from the photon case where there is only a $J=1$ component, for both on-shell and off-shell photons. The possible neutrino final states (Majorana or Dirac) for the graviton case are :
 \be
\begin{split}
J=0~:~~&S=0~: ~~^{1}S_{0}~~~,\\
&S=1~:~~^{3}P_{0}~~~,\\
\\
J=2~:~~&S=0~:~~^{1}D_{2}~~~,\\
&S=1~:~~^{3}P_{2},~^{3}D_{2},~^{3}F_{2}~~~,\\
\end{split}
\label{angstates}
\eq
where as usual $J=L+S$ and we have used the standard spectroscopic notation $^{2S+1}L_{J}$ for the angular momentum states. If we take the on-shell limit for the graviton some of the states cancel to yield the expected $J_{3}=\pm \,2$ helicity~\cite{ddsw}. Finally, note that the six angular momentum states of Eq.\,(\ref{angstates}) are related to the six operators that appear in the Dirac neutrino-graviton vertex. This association will be explicitly derived in the non-relativistic limit.
\par
Let us now consider the Majorana case where the two final state particles are identical fermions. We require this final state to be anti-symmetric under the exchange $[p1,s1]_{i}\leftrightarrow[p2,s2]_{i}$. For an identical (Majorana) two-particle final state we have under the permutation operator $\mathbb{P}_{12}\left([p1,s1]_{i}\leftrightarrow[p2,s2]_{i}\right)$ :
\be
\mathbb{P}^{M}_{12}\left(^{2S+1}L_{J}\right)~=~\left(-1\right)^{L+S+1}~~~.
\eq
Using the above expression for the Majorana final states we find that five of the final states are antisymmetric except $D^{3}_{2}$ which has
\be
\mathbb{P}^{M}_{12}\left(^{3}D_{2}\right)~=~+1~~~.
\label{fdam}
\eq
This again implies that in the Majorana neutrino case one can have only \textit{five} graviton form factors to be consistent with Fermi-Dirac statistics. When we derive the angular momentum correspondence of the graviton-vertex operators explicitly in the non-relativistic limit, we will show concretely  that Eq. (\ref{fdam}) indeed implies Eq. (\ref{gm3}). 
\par
Now we will try to associate each of the form factors (in the non-relativistic limit) with states of fixed  angular momentum. The off-shell graviton polarization tensor $\epsilon_{\mu\nu}[J, J_{3}]$ with $J=0$ and $J=2$ may be written in terms of the $j=1$, $j^{'}=1$ polarization vectors $\eta_{\mu}[j,j_{3}]$, $\eta_{\mu}[j^{'},j^{'}_{3}]$ of the spin-1 field (photon) as
\be
\epsilon_{\mu\nu}[J, J_{3}]=\sum_{j_{3}+j^{'}_{3}=J_{3}}\mathcal{C}(J\,1\,1;J_{3}\,j_{3}\,j^{'}_{3})\,\eta_{\mu}[1,j_{3}]\,\eta_{\nu}[1,j^{'}_{3}]~.
\label{gpolt}
\eq
Here $\mathcal{C}(J\,j\,j^{'};J_{3}\,j_{3}\,j^{'}_{3})$ are the usual Clebsch-Gordan coefficients. We will work in the rest frame of the off-shell graviton $h^{*}$ so as to simplify the analysis.
The photon polarization vectors of good angular momentum in this special frame  may be chosen to be
\be
\begin{split}
\eta[1,+1]&=\frac{1}{\sqrt{2}}\left(0,-1,-i,0\right)~~~,\\
\eta[1,0]&=~~~\left(0,0,0,1\right)~~~,\\
\eta[1,-1]&=\frac{1}{\sqrt{2}}\left(0,1,-i,0\right)~~~.\\
\end{split}
\eq
 In this case the five graviton polarization tensors with $J=2$ may be written explicitly using the above formula as
\bea
\nn
\epsilon[2,\pm 2]&=~~~~\frac{1}{2}
\begin{pmatrix}
0&0&0&0 \\
0&1&\pm i&0 \\
0&\pm i&-1&0 \\
0&0&0&0 \\
\end{pmatrix}~~~,\\ \nn
\epsilon[2,\pm 1]&=\frac{1}{2}
\begin{pmatrix}
0&0&0&0 \\
0&0&0&\mp 1 \\
0&0&0&-i \\
0&\mp 1&-i&0 \\
\end{pmatrix}~~~,\\
\epsilon[2,0]&=~~\frac{1}{\sqrt{6}}
\begin{pmatrix}
0&0&0&0 \\
0&-1&0&0 \\
0&0&-1&0 \\
0&0&0&2 \\
\end{pmatrix}~~~,
\label{gpol2}
\eea
and the $J=0$ polarization tensor is
\bea
\epsilon[0,0]=~~\frac{1}{\sqrt{3}}
\begin{pmatrix}
0&0&0&0 \\
0&-1&0&0 \\
0&0&-1&0 \\
0&0&0&-1 \\
\end{pmatrix}~~~.
\label{gpol0}
\eea
In the rest frame of the off-shell graviton we have the neutrino momenta $\vec{p}_{2}=-\vec{p}_{1}=\vec{p}$ and without loss of generality we normalize it to have unit magnitude. Following the notation in \cite{kay} we label $\phi^{\dag}_{2}\phi^{c}_{1}=S$ and $\phi^{\dag}_{2}\vec{\sigma}\phi^{c}_{1}=\vec{T}$ for the singlet and triplet states. Here $\phi_{1}$ and $\phi_{2} $ are again the large component Pauli spinors of the two neutrinos. With these definitions, for the states of interest, we find that the angular momentum eigenfunctions take the form
\bea
\nn
^{1}S_{0}&:&~S~~~,\\ \nn
^{3}P_{0}&:&~\hat{p}\cdot\vec{T}~~~,\\ \nn
^{1}D_{2}&:&~\left(3~\hat{p}\otimes\hat{p}-1\otimes 1\right)~S~~~,\\ \nn
^{3}P_{2}&:&~\hat{p}\otimes\vec{T},~(\hat{p}\cdot\vec{T})~1\otimes 1~~~,\\ \nn
^{3}D_{2}&:&~\hat{p}~\otimes~(\vec{T}\times\hat{p})~~~,\\
^{3}F_{2}&:&~5~(\hat{p}\cdot\vec{T})~\hat{p}\otimes\hat{p}+(\hat{p}\cdot\vec{T}) ~1\otimes 1-\hat{p}\otimes\vec{T}~.
\label{amef}
\eea
Here $\otimes$ denotes a direct product of vectors and symmetrization is implicit. Working in a non-chiral Dirac basis one can take the non-relativistic limit of the operators in Eq.\,(\ref{fullforms}) in the standard way and express them in terms of the angular momentum eigenfunctions above. Using Eqs.\,(\ref{gpolt})- (\ref{amef}) along with the properties of the polarization tensors gives
\bea
\nn
&&\epsilon^{\mu\nu}[0,J_{3}]\cdot\bar{u}_{2}\Big(q^{2}g_{\mu\nu}-q_{\mu}q_{\nu}\Big)v_{1}\xrightarrow{NR}-\frac{\sqrt{3} q^{2}}{m_{\nu}}\hat{p}\cdot\vec{T}\\ \nn
\\ \nn
&&\epsilon^{\mu\nu}[2,J_{3}]\cdot\bar{u}_{2}\Big(\gamma_{\mu}r_{\nu}+\gamma_{\nu}r_{\mu}\Big)v_{1}\xrightarrow{NR}\frac{2\epsilon^{ij}}{5m_{\nu} ^{2}}\Big[5(\hat{p}\cdot\vec{T})\hat{p}_{i}\hat{p}_{j}\\ \nn
&&+ (\hat{p}\cdot\vec{T}) \delta_{ij}-(\hat{p}_{\{i}\vec{T}_{j\}})\Big]-\frac{\epsilon^{ij}}{10 m_{\nu} ^{2}}\Big[4(\hat{p}\cdot\vec{T})\delta_{ij}+(\hat{p}_{\{i}\vec{T}_{j\}})\Big]\\ \nn
&&\epsilon^{\mu\nu}[0,J_{3}]\cdot\bar{u}_{2}\Big(\gamma_{\mu}r_{\nu}+\gamma_{\nu}r_{\mu}\Big)v_{1}\xrightarrow{NR} \frac{\hat{p}\cdot\vec{T}}{m^{2}_{\nu}}\\ \nn
\\ \nn
&&\epsilon^{\mu\nu}[2,J_{3}]\cdot\bar{u}_{2}(r_{\mu}r_{\nu})v_{1}\xrightarrow{NR}\frac{4\epsilon^{ij}}{5m_{\nu} }\Big[5(\hat{p}\cdot\vec{T})\hat{p}_{i}\hat{p}_{j} \\ \nn
&&+(\hat{p}\cdot\vec{T})\delta_{ij}-(\hat{p}_{\{i}\vec{T}_{j\}})\Big]+\frac{4\epsilon^{ij}}{5 m_{\nu}}\Big[(\hat{p}_{\{i}\vec{T}_{j\}})-(\hat{p}\cdot\vec{T})\delta_{ij}\Big] \\ 
&&\epsilon^{\mu\nu}[0,J_{3}]\cdot\bar{u}_{2}(r_{\mu}r_{\nu})v_{1}\xrightarrow{NR}\frac{4\hat{p}\cdot\vec{T}}{m_{\nu}}
\label{amnrl1}
\eea

\bea
 \nn
\epsilon^{\mu\nu}[0,J_{3}]\cdot\bar{u}_{2}\left(q^{2}g_{\mu\nu}-q_{\mu}q_{\nu}\right)\gamma^{5}v_{1}&&\xrightarrow{NR}\\ \nn
&&\sqrt{3} q^{2}\left(1+\frac{p^{2}}{4m_{\nu} ^{2}}\right)S\\ \nn
\\ \nn
\epsilon^{\mu\nu}[2,J_{3}]\cdot\bar{u}_{2}\left(\gamma_{\mu}r_{\nu}+\gamma_{\nu}r_{\mu}\right)\gamma^{5}v_{1}&&\xrightarrow{NR}\frac{2\epsilon^{ij}}{m_{\nu}}\left[\hat{p}_{\{i}(\hat{p}\times\vec{T})_{j\}}\right]\\ \nn
\\ \nn
\epsilon^{\mu\nu}[2,J_{3}]\cdot\bar{u}_{2}(r_{\mu}r_{\nu})\gamma^{5}v_{1}&&\xrightarrow{NR} -\frac{4}{3}\left(1+\frac{p^{2}}{4m_{\nu}^{2}}\right)\\ \nn
&& \epsilon^{ij}\left(3\hat{p}_{i}\hat{p}_{j}-\delta_{ij}\right)S~~~\\ \nn
\epsilon^{\mu\nu}[0,J_{3}]\cdot\bar{u}_{2}(r_{\mu}r_{\nu})\gamma^{5}v_{1}&&\xrightarrow{NR}\\ 
&& 4 \left(1+\frac{p^{2}}{4m_{\nu}^{2}}\right)S \label{amnrl2}
\eea
where, again $\{\}$ denotes complete symmetrization with respect to the relevant indices. 
\par
Comparing Eqs.\,(\ref{amef}),\,(\ref{amnrl1}) and\,(\ref{amnrl2}) we deduce the angular momentum associations, in the non-relativistic limit, for the graviton vertex operators 
\bea
\label{ffanga}
q^{2}g_{\mu\nu}-q_{\mu}q_{\nu}&:&~^{3}P_{0}~~~,\\ \nn
(q^{2}g_{\mu\nu}-q_{\mu}q_{\nu})\gamma_{5}&:&~^{1}S_{0}~~~,\\ \nn
\gamma_{\mu}r_{\nu}+\gamma_{\nu}r_{\mu}&:&~^{3}F_{2} \oplus\,^{3}P_{2}~;~ ^{3}P_{0}~~~,\\ \nn
(\gamma_{\mu}r_{\nu}+\gamma_{\nu}r_{\mu})\gamma_{5}&:&~^{3}D_{2}~~~,\\ \nn
r_{\mu}r_{\nu}~~~~~~&:&~^{3}F_{2}\oplus\, ^{3}P_{2}~;~ ^{3}P_{0}~~~,\\ \nn
(r_{\mu}r_{\nu})\gamma_{5}~~~&:&~^{1}D_{2}~;~^{1}S_{0}~~~.\nn
\eea
This correspondence of the vertex operators with the angular momentum states in the non-relativistic limit again provides an interpretation for the associated form factors and is complementary to the earlier results we derived in Eqs.\,(\ref{zcpl}) and (\ref{aved}). 
\par
Furthermore we see from Eq.\,(\ref{ffanga}) that the $^{3}D_{2}$ state is indeed associated with the operator $(\gamma_{\mu}r_{\nu}+\gamma_{\nu}r_{\mu})\gamma^{5}$ and hence the form factor $G_{3}(q^{2})$. This relates the previous results for Majorana neutrinos in Eqs.\,(\ref{gm3}) and (\ref{fdam}), and clearly shows that they are indeed consistent with each other.
\par
Let us now consider matrix element calculations from the viewpoint of Feynman diagrams in the Dirac and Majorana cases. An important point to note is that for Majorana neutrinos coupled to a graviton there are additional diagrams that have to be included in the calculation of $\langle\nu( p^{'})_{i}\lvert\mathcal{J}_{\mu\nu}\rvert\nu(p)_{i}\rangle$. To be more specific we mean that for Majorana neutrinos there would be an additional ``charge conjugated" diagram corresponding to each Feynman diagram of the Dirac case. Based on this observation we can try to understand how the matrix elements in the Majorana case may be calculated from the Dirac neutrino calculation. Again, we modify an argument used in the neutrino-photon case~\cite{kay}. Let us assume that for Dirac neutrinos the t-channel matrix element at one-loop is
\be
\langle\nu_{i}( p^{'}, s^{'})\lvert\Delta\hat{\mathcal{J}}^{D}_{\alpha\beta}\rvert~\nu_{i}(p,s)\rangle_{\text{\tiny{1-loop}}}=\kappa~\Gamma_{\alpha\beta}^{D}(g,\gamma_{5},p,q)
\label{dfeyd}
\eq
where $q= p^{'}-p$ and $g$ is the electroweak coupling constant. For the Majorana case the charge conjugate diagram will have all particles replaced by their charge conjugates and the electroweak projection operator $P_{L}\rightarrow P_{R}$. Unlike in the electromagnetic case (with coupling constant $e$) the coupling constant in the present case, namely $\kappa$, does not change sign under charge conjugation and hence for the corresponding Majorana case we have at one-loop
\bea
\label{mfeyd}
\langle\nu_{i}( p^{'}, s^{'})\lvert\Delta\hat{\mathcal{J}}^{M}_{\alpha\beta}\rvert~\nu_{i}(p,s)\rangle_{\text{\tiny{1-loop}}}&=&\kappa~\Gamma_{\alpha\beta}^{D}(g,\gamma_{5},p,q)\\ \nn
&+&\kappa~\Gamma_{\alpha\beta}^{D}(g,-\gamma_{5},p,q)~.\nn
\eea
This suggests that at 1-loop, \emph{compared to the Dirac neutrino, the vertex factors for the Majorana neutrino will \textit{not have} any terms proportional to $\gamma_{5}$}. This conclusion is opposite to the equivalent conclusion in the neutrino-photon case~\cite{kay}. 
\par
It may be puzzling at first that this line of reasoning suggests in addition to the vanishing of $G^{M}_{3}(q^{2})$ (as in Eq. (\ref{gm3}))  that the form factors 
\ba
G^{M}_{1}(q^{2})~=~G^{M}_{2}(q^{2})~=~0~~~.
\ea
We would have naively expected both the argument from Fermi-Dirac statistics as well as the above to lead to identical conclusions. Let us try to understand more thoroughly the reason for this difference. We note that the 1-loop argument above relied on the charge conjugation property of the Majorana neutrino
\ba
\Psi_{M}=\Psi^{\mathcal{C}}_{M}~~~~~~~\text{(Majorana Condition)}~~~,
\ea
and an implicit assumption that there were \textit{no CP phases} at 1-loop to start with in the graviton-Dirac neutrino 1-loop diagram. Thus it seems pertinent to explore the consequences of exact (or approximate) CP invariance on the final states, in the $s$-channel process, to understand the relation between the two arguments.
\par
For Dirac particles in the final state, of the s-channel process $\langle\nu( p^{'})_{i}\nu(p)_{i}\lvert\mathcal{J}_{\mu\nu}\rvert0\rangle$, we have the well-known result
\ba
\zeta^{\text{\tiny{D}}}_{\text{\tiny{CP}}}\left(^{2S+1}L_{J}\right)~=~\left(-1\right)^{S+1}~~~.
\ea
For clarity we have briefly reviewed the $C$, $P$ and $CP$ properties of Dirac and Majorana two-particle final states in Appendix A. Comparing the above expression with the angular momentum states in Eq. (\ref{angstates}) we find that all Dirac neutrino final states are CP even except 
\ba
\zeta^{\text{\tiny{D}}}_{\text{\tiny{CP}}}\left(^{1}S_{0}\right)~=~\zeta^{\text{\tiny{D}}}_{\text{\tiny{CP}}}\left(^{1}D_{2}\right)~=~-1~~~.
\ea
\par
For Majorana neutrino final states the CP parity is given by
\ba
\zeta^{\text{\tiny{M}}}_{\text{\tiny{CP}}}\left(^{2S+1}L_{J}\right)~=~\left(-1\right)^{L+1}~~~,
\ea
which is also the expression for the usual parity (ie. spatial reflection) since charge conjugation is trivial for the Majorana final states. When this is applied to the possible final angular momentum states of the Majorana neutrinos we find that all states are CP even except
\ba
\zeta^{\text{\tiny{M}}}_{\text{\tiny{CP}}}\left(^{1}S_{0}\right)~=~\zeta^{\text{\tiny{M}}}_{\text{\tiny{CP}}}\left(^{1}D_{2}\right)~=~\zeta^{\text{\tiny{M}}}_{\text{\tiny{CP}}}\left(^{3}D_{2}\right)~=~-1~~~.
\ea
\par
The $^{3}D_{2}$ Majorana state is already ruled out from arguments of Fermi-Dirac statistics. So, if it is the case that in the lepton sector CP invariance holds exactly, or the CP phases are negligible for all practical considerations, then from the associations in Eq. (\ref{ffanga})
\ba
G_{1}(q^{2}),~G_{2}(q^{2})~&:&~^{1}S_{0},~^{1}D_{2}~~~,\\
G_{3}(q^{2})~&:&~^{3}D_{2}~~~,\\
\ea
 we obtain, to all orders in perturbation theory, in addition to Eq. (\ref{gm3}) 
 \ba
G^{D}_{1}(q^{2})&\xrightarrow{CP}~&0~~~~~G^{D}_{2}(q^{2})\xrightarrow{CP}~0~~~,\\ \nn
G^{M}_{1}(q^{2})&\xrightarrow{CP}~&0~~~~~G^{M}_{2}(q^{2})\xrightarrow{CP}~0~~~. \nn
\ea
This is an interesting observation in the context of whether there is CP violation at all in the lepton sector, especially since in the Majorana case there could be additional Majorana phases in the electroweak mixing matrix. One of the consequences of the above result is that \emph{there is a definite relation between the presence of CP phases (Dirac or Majorana) in the lepton sector and the vanishing of \textit{two} of the graviton-neutrino form factors}.
\par
Thus it may be speculated that irrespective of whether neutrinos are Dirac or Majorana, \textit{if it is feasible to probe the graviton form factors $G_{1}(q^{2})$ and $G_{2}(q^{2})$ in future neutrino experiments, it could potentially yield information about CP phases in the charged lepton-neutrino sector}. 
But note that $G_{2}(q^{2})$, similar to $F_{2}(q^{2})$ is associated with a chirality flipping operator and hence suppressed by $\mathcal{O}(m_{\nu}/E_{\nu})$. Thus the preliminary indications are that it might be challenging to probe $G_{2}(q^2)$ in experiments, unless one considers very low energy neutrino sources, to speculate, say maybe from the cosmic neutrino background (cf.~\cite{astroneu}).
\par
Now let us consider the expressions in Eqs. (\ref{Eforms}) and (\ref{Dforms}). In the Majorana off-diagonal case $i\neq j$, for the s-channel, we would have (See Appendix A)
\be
\zeta^{\text{\tiny{M}}}_{\text{\tiny{CP}}} \left(^{2S+1}L_{J}\right) =\eta^{*}_{i}\eta^{*}_{j}(-1)^{L}~~~.
\eq
Thus there are two possibilities for the s-channel final state intrinsic parity, $\eta^{*}_{i}\eta^{*}_{j}=-1$ and $\eta^{*}_{i}\eta^{*}_{j}=+1$. Since the product of intrinsic parities in the t-channel $\langle\nu( p^{'})_{j}\lvert\mathcal{J}_{\mu\nu}\rvert\nu(p)_{i}\rangle$ is opposite to the product of intrinsic parities in the equivalent s-channel $\langle\nu( p^{'})_{j}\nu(p)_{i}\lvert\mathcal{J}_{\mu\nu}\rvert0\rangle$, the two cases of $\eta^{*}_{i}\eta^{*}_{j}=-1,\,+1$  would correspond, in the t-channel, to the vanishing of $(E_{1},~E_{2},~E_{3})$ and $(D_{1},~D_{2},~D_{3})$ respectively under the approximations we have made. This result, argued from the CP symmetries of the angular momentum states, is again consistent with the equivalent result in~\cite{klak}.
\par
We observe specifically from Eqs. (\ref{gm3}), (\ref{dfeyd})  and (\ref{mfeyd}) that the graviton vertex matrix elements for Dirac and Majorana cases seem to generically come out different. We will now show that irrespective of this apparent difference the \textit{practical Majorana-Dirac confusion theorems} ~\cite{kay} first discussed in the context of neutrino scattering and electromagnetic interactions still hold in the gravitational case and render any difference undetectable. The main reasons for this conclusion, as pointed out long back, are that the electroweak currents are left handed and that the neutrino masses are tiny~\cite{kay}. We will see in what follows that the arguments in the electromagnetic case may be essentially carried over to the gravitational case. 
\par
We note that the crucial difference in the matrix elements is that in the Majorana case we have just $F^{M}_{3}(q^{2})$ associated with the operator $\gamma_{\{\mu}r_{\nu\}}$ while in the Dirac case we have both 
$F^{D}_{3}(q^{2})$ and $G^{D}_{3}(q^{2})$. Thus as in the analogous photon vertex case if it so happens that
\be
\label{dmct}
\frac{F^{M}_{3}(q^{2})}{4}\simeq\frac{F^{D}_{3}(q^{2})}{4}+ q^{2}G^{D}_{3}(q^{2})
\eq
to say $\mathcal{O}(m_{\nu}/E_{\nu})$, then experimentally one would never be able to distinguish Majorana neutrinos from Dirac neutrinos interacting with gravity. We will in fact see that this is the case. Let $L$ and $R$ denote left and right chiralities. $P_{L}$ and $P_{R}$ will denote the electroweak left and right projection operators.
\par
At tree-level we must have
\be
\label{tlr}
\frac{F^{M,0}_{3}}{4}=\frac{F^{D,0}_{3}}{4}+ G^{D,0}_{3}
\eq
where $F^{M,0}_{3}$, $F^{D,0}_{3}$ and $G^{D,0}_{3}$ are the terms generated at tree-level. This is because, crudely speaking, at tree level there are no additional Feynman diagrams for the Majorana neutrino compared to the Dirac neutrino. 
\par
Now, we know that a mass term in the Dirac equation mixes $L$ and $R$ chiralities. In any electroweak loop diagram there are left projection operators sitting at the initial and final electroweak vertices. So, in the limit of vanishing neutrino masses, the electroweak left projection operators attenuate any diagram with an incoming or outgoing $R$ neutrino.
Thus beyond tree-level we must have
\be
\langle\nu_{R}\lvert \Delta\mathcal{J}_{\mu\nu}\rvert\nu_{R}\rangle\xrightarrow{m_{\nu}\rightarrow0} 0
\eq
owing to the left-handed nature of the weak currents and the extremely small neutrino mass. The above expression and Eq. (\ref{fullforms}) therefore imply
\be
\label{ilr}
\frac{\Delta F^{D}_{3}(q^{2})}{4}\simeq q^{2}\Delta G^{D}_{3}(q^{2})
\eq
where we have used the result
\ba
\gamma_{5} u^{\nu}_{R}\simeq- u^{\nu}_{R}~,~~\gamma_{5} u^{\nu}_{L}\simeq+ u^{\nu}_{L} ~~~.
\ea
Also, we have from Eq. (\ref{mfeyd}) the observation
\bea
\nn
&&\bar{u}_{L}\frac{\Delta F^{M}_{3}}{4}u_{L}=\bar{u}_{L}\left[\frac{\Delta F^{D}_{3}}{4}+q^{2} \Delta G^{D}_{3}\right]u_{L}+\\ 
&&\bar{u}_{L}\left[\frac{\Delta F^{D}_{3}}{4}- q^{2}\Delta G^{D}_{3}\right]u_{L}=\bar{u}_{L}\left[\frac{\Delta F^{D}_{3}}{2}\right]u_{L}~.
\label{lhr}
\eea
This is because, as mentioned previously, for Majorana neutrinos there is an additional ``charge conjugated" diagram, with $P_{L}\rightarrow P_{R}$ and $l^{-}\rightarrow l^{+}$, corresponding to each Feynman diagram of the Dirac case.
\par
Putting together the results (\ref{tlr})-(\ref{lhr}) we have
\bea
\nn
&&\frac{F^{D}_{3}}{4}+ q^{2}G^{D}_{3}=\frac{F^{D,0}_{3}}{4}+ G^{D,0}_{3}+\frac{\Delta F^{D}_{3}}{4}\\ \nn
&&+ q^{2}\Delta G^{D}_{3}=\frac{F^{M,0}_{3}}{4} +\frac{\Delta F^{D}_{3}}{4}+ q^{2}\Delta G^{D}_{3}\\ \nn
&&\simeq\frac{F^{M,0}_{3}}{4} +\frac{\Delta F^{D}_{3}}{4}+ \frac{\Delta F^{D}_{3}}{4}=\frac{F^{M,0}_{3}}{4} +\frac{\Delta F^{D}_{3}}{2} \\ \nn
&&\simeq\frac{F^{M,0}_{3}}{4} +\frac{\Delta F^{M}_{3}(q^{2})}{4}=\frac{F^{M}_{3}}{4}~~~. 
\eea
Thus Eq.\,(\ref{dmct}) is indeed satisfied for the neutrino-graviton vertex. So, \emph{in spite of theoretical differences in the graviton vertex of Majorana and Dirac neutrinos, to $\mathcal{O}(m_{\nu}/E_{\nu})$, the two cases would be practically indistinguishable in any weak-field space-time geometry}.
\par
We note that this is in disagreement with some of the recent claims in the literature by D. Singh \textit{et. al}~\cite{dng} based on studies in the gravitational phase formalism~\cite{gphase}. Also see J. F. Nieves and P. B. Pal ~\cite{dng} for a similar discussion that gives the answer in the negative and some related comments in~\cite{ggibbons}.
\par
Finally, it is to be pointed out that most of our analyses for the graviton-neutrino interactions so far have been fairly general, based only on symmetry principles. Thus, we believe that most of the conclusions derived so far are equally applicable to other Majorana fermions that might exist in nature and couple to gravity. For instance, in the context of the proposed \textit{Minimal Supersymmetric Standard Model} (MSSM)~\cite{psch} there are additional Majorana fermions such as the gluino ($\tilde{g}$) and the neutralino ($\tilde{\chi}^{0}$). In many models the $\tilde{\chi}^{0}$ plays the role of the lightest supersymmetric partner and may be a large component of the cold dark matter (CDM) in the universe. So if one is interested in the interaction
\be
\tilde{\chi}^{0}\xrightarrow{~h^{*}~}\tilde{\chi}^{0}
\eq
in the presence of weak gravitational fields then all our conclusions may be carried over without almost any change for the Majorana neutralino $\tilde{\chi}^{0}$.
\section{Graviton-Neutrino coupling at 1-loop} 
\par
We now perform a calculation of the mass off-diagonal neutrino-graviton form factors at 1-loop, to lowest order under the assumption that the neutrino energy $E_{\nu}\gg m_{\nu}$. From this calculation we will see how the gravitational neutrino transition form factors vary with $q^{2}$ and also be able to estimate an $ i\rightarrow j$ neutrino transition mass radius. We are mainly interested in the mass off-diagonal case ($i\neq j$) because if it is non-vanishing it may be thought of as a graviton mediated ``penguin" transition $\nu_{i}\rightarrow\nu_{j}$ for non-zero $q^{2}$. Such an $i\rightarrow j$ transition would be a unique quantum mechanical effect with no equivalent in the linearized classical theory.
\par
In the presence of weak gravitational fields, it is well known that we may use a perturbative approach to calculate the effects of the gravitational field on particles (see~\cite{qftcs} and references therein). Specifically, from the principle of general covariance we may derive the Feynman rules for the graviton-fermion and graviton-massive vector boson interactions. This is achieved by making the associations
\ba
\int~d^{4}x~\mathcal{L}_{\text{\tiny{SM}}}\big[\eta_{\mu\nu}\big]&\rightarrow&~\int~d^{4}x~\sqrt{-g}~\mathcal{L}_{\text{\tiny{SM}}}\big[g_{\mu\nu}\big]\\ \nn
\big(\vec{\nabla}_a\big)_{\Psi_{\textit{f}}}&~\rightarrow&~e^{\alpha}_{a}\left(\partial_{\alpha}+iw_{\alpha}\right)
\ea
in the action integral and expanding all the relevant quantities around the flat metric using Eq.\,(\ref{linmet})~\cite{qftcs} . In the above, $w_{\alpha}$ are the field connections which are expressed in terms of the vierbeins/tetrads $e^{\alpha}_{a}$~\cite{qftcs}. 
\par
This effective field theory approach is the one we will follow to study the potential effects of weak gravitational fields on the transition scattering of high energy neutrinos. The relevant graviton Feynman rules for our calculation are listed in Appendix B. The graviton Feynman rules we derive for the standard model follow the conventions of Cheng and Li~\cite{psch}. The rules we use seem to be in slight disagreement with the Feynman rules used in ~\cite{pbpjn} (specifically regarding some of the relative signs of the terms), but are consistent with the rules derived by S.Y. Choi \textit{et. al.}~\cite{qftcs}. 
\par
The 1-loop diagrams we are interested in for the $\nu_{i} \neq \nu_{j}$ case, to lowest order, are shown in Figs.\,(\ref{4g1}) and (\ref{gwc}). For the transition form factor ($E_{k}(q^{2}),~ D_{k}(q^{2})$) calculation these are the only Feynman diagrams we consider. If we are instead interested in the diagonal form factors ($F_{k}(q^{2}),~ G_{k}(q^{2})$) there would also be additional Feynman diagrams involving the $Z$ boson. We choose to work in the \textit{Feynman-`t Hooft gauge} using dimensional regularization and will adopt the $\overline{MS}$ scheme. Since we are working to lowest order we neglect terms that are proportional to neutrino masses, since by assumption $E_{\nu}\gg m_{\nu}$. Also, we do not include diagrams with ghost fields running in the loop which are expected to give corrections at a higher order. These are the approximations in our loop calculations.
\par
Our task now is to explicitly evaluate the diagrams in the presence of a weak gravitational field to lowest order, obtain the $q^{2}$ dependence of the gravitational form factors and then estimate the neutrino transition mass radius. Let us take up the first diagram shown in Fig. (\ref{4g1}). It has a topology that is unique to the graviton interaction and is not present in the photon case. As before we denote the neutrino mass-eigenstates by the latin indices $i,j$ and $l$ denotes the lepton flavor. The Pontecorvo-Maki-Nakagawa-Sakata (PMNS) matrix~\cite{mns} is denoted by $U_{li}$. The $W^{\pm}$ mass and the mass of the charged lepton $l^{-}$ are denoted by $M_{W}$ and $m_{l}$ respectively. $g$ is the electroweak coupling constant.
\par
Using the Feynman rules for linearized gravity, as reviewed in Appendix B, the spin-2 gauge current for the first Feynman diagram with $i\neq j$ in Fig.\,(\ref{4g1}) is
\ba
&&\bar{u}_{j}i\Delta\mathcal{J}_{\mu\nu}^{(1)} u_{i}\sim\sum_{l=e,\mu,\tau} \int \frac{d^4 k}{(2\pi)^4} \bar{u}_{j}(p') 
\big(\frac{ig}{\sqrt{2}} \gamma_{\sigma} P_L U_{lj}^* \big) \\ \nn
&& \big(i \frac{\slashed{k} + m_l}{k^2 - m_l^2} \big)  \big(i \frac{\kappa g}{2\sqrt{2}} \Gamma_{\mu \nu \rho \lambda} \gamma^{\lambda} 
P_L U_{li} \big) u_{i}(p)\\ 
&& \big(-i \frac{\eta^{\rho \sigma}}{
(k-p')^2 - M_W^2} \big) \\
&=& \frac{\kappa g^2 U_{lj}^* U_{li}}{4}\bar{u}_{j}(p')\\ \nn
&&\bigg\{ \int \frac{d^4 k}{(2\pi)^4}
\frac{\gamma^{\rho} P_L (\slashed{k} + m_l) \Gamma_{\mu \nu \rho 
\lambda} \gamma^{\lambda} P_L }{(k^2 - m_l^2)((k-p')^2 - M_W^2)}\bigg\}~ u_{i}(p)~~,
\ea
where 
\ba
\Gamma_{\mu \nu \rho \lambda}  &=& \eta_{\mu \nu} \eta_{\rho 
\lambda} -\frac{1}{2} \left(\eta_{\mu \rho} \eta_{\nu \lambda} + \eta_{\mu 
\lambda} \eta_{\nu \rho} \right)~~.
\ea
\par
For the similar Feynman diagram in Fig.\,(\ref{4g1}), where the graviton is coupled to the final vertex we have
\ba
&&\bar{u}_{j}i\Delta\mathcal{J}_{\mu\nu}^{(2)}u_{i}\sim\sum_{l=e,\mu,\tau} \int \frac{d^4 k}{(2\pi)^4}\bar{u}_{j}(p') 
i \frac{\kappa g}{2\sqrt{2}} \Gamma_{\mu \nu \rho \lambda} \gamma^{\lambda} 
P_L U^{*}_{lj} 
\\ \nn
&&\big( i \frac{\slashed{k} + m_l}{k^2 - m_l^2} \big) \big(\frac{ig}{\sqrt{2}} \gamma_{\sigma} P_L U_{li} \big)
 u_{i}(p)\big(-i \frac{\eta^{\rho \sigma}}{
(k-p)^2 - M_W^2} \big) \\
&=& \frac{\kappa g^2 U_{lj}^* U_{li}}{4} \bar{u} _{j}(p')\\ \nn
&&\bigg\{ \int \frac{d^4 k}{(2\pi)^4} \frac{\Gamma_{\mu \nu \rho \lambda} \gamma^{\lambda}  P_L (\slashed{k} +
m_l)  \gamma^{\rho} P_L}{(k^2 - m_l^2)((k-p)^2 - M_W^2)}\bigg\} u(p)~~~.
\nonumber 
\ea
\begin{figure}
\begin{center}
\includegraphics[width=7cm,angle=0]{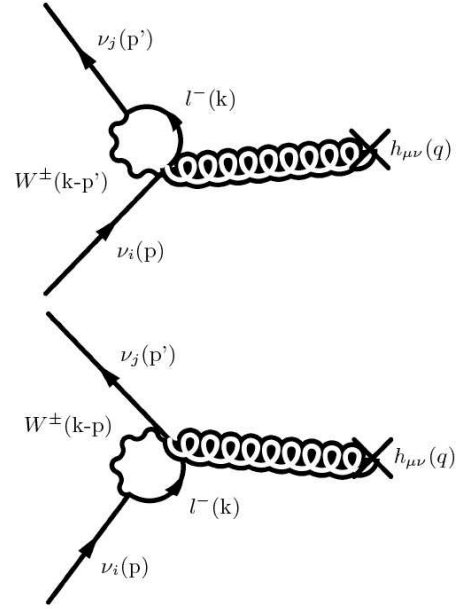}
\end{center}
\caption{Four-point neutrino graviton diagrams at 1-loop with $i \neq j$.}
\label{4g1}
\end{figure}
\par
The above spin-2 gauge currents for the diagrams in Fig.\,(\ref{4g1}) may be simplified using the standard techniques of Dirac algebra, assuming $E_{\nu}\gg m_{\nu}$. This gives finally the approximate expression for the gauge current
\bea
\label{4ptc}
&&\bar{u}_{j}~i\Delta\mathcal{J}_{\mu\nu}^{(\text{4 pt.})}u_{i} \sim\sum_{l}\frac{i\kappa g^{2}}{32\pi^{2}} U_{lj}^{*} U_{li}\int_{0}^{1} dx\\ \nn
&&\bigg[x\log\Delta(M_{W}^{2},m_{l}^{2})\bigg]  \bar{u}_{j}\gamma_{\{\mu}r_{\nu\}}\hat{P}_L u_{i}
\eea
where
\ba
\Delta(M_{W}^{2},m_{l}^{2})=xM_{W}^{2}+(1-x)m_{l}^{2}~~~.
\ea
\par
Studying the above 1-loop expression we see that indeed we do get an operator form that is consistent with our expectations of gauge invariance. The only form factors that are generated at this order are $E_{3}(q^{2})$ and $D_{3}(q^{2})$. It is noted that none of the other form factors are generated at this order from these diagrams. As previously mentioned, the form factors of the type $E_{2}(q^{2})$ and $D_{2}(q^{2})$ corresponding to $r_{\alpha}r_{\beta}$ mediate chirality flips for the incoming neutrino and are thereby further suppressed by $\mathcal{O}(m_{\nu}/E_{\nu})$. Since our approximate calculation neglects terms proportional to $m_{\nu}$ these terms are not retained.
\par
Also note that the contributions from the four-point graviton diagrams do not have a $q^{2}$ dependence at this order of approximation. As we shall see, this implies that when we enforce correct renormalization for the spin-2 currents they would not contribute. Thus this would imply that they do not contribute to the $\nu_{i}\rightarrow \nu_{j}$ transition mass radius of the neutrino as well at leading order.
\par
The last thing we would like to point out is that the Glashow-Iliopoulos-Maiani (GIM) mechanism~\cite{gim} is again relevant here. Note from the above that in the limit of degenerate charged lepton masses, the loop result vanishes for $i\neq j$.
\par
The next diagram we consider is the one where the graviton is coupled to the $W^{\pm}$ gauge boson in the loop. In the equivalent photon case, such a diagram would only have contributed to next order in $1/M^{2}_{W}$. A quick look at the Feynman rule for the gauge boson-graviton coupling would convince us that this is not true in the graviton case. This is because in the graviton case the numerator of the gauge boson-graviton coupling has a term that is proportional to $M^{2}_{W}$.
\par
For the graviton coupling to the W boson we have 
\ba
&&\bar{u}_{j}i \Delta J_{\mu \nu}^{(3)} u_{i}\sim\sum_{l=e,\mu,\tau}\int \frac{d^4k}{(2\pi)^4} \frac{\kappa g^2 U_{lj}^* 
U_{li}}{2}\big(\eta^{\alpha \rho} ~\eta^{\lambda \beta}\big) \bar{u}_{j}(p')  \nonumber \\
&&\frac{\big[\gamma_{\beta} P_L \Gamma^{'}_{\mu \nu \rho \lambda} 
(-\slashed{k} + m_l) \gamma_{\alpha} P_L \big]u_{i}(p) }{(p+k)^2-M_W^2
(p+k')^2-M_W^2 (k^2-m_l^2)} \\ \nn
\ea
where 
\ba
&&\Gamma^{'}_{\mu \nu \rho \lambda}= 
 - \frac{M^{2}_{W}}{2}\big[\eta_{\mu\nu}\eta_{\alpha\beta}
-\eta_{\mu\alpha}\eta_{\nu\beta}-\eta_{\nu\alpha}\eta_{\mu\beta}\big]\\ \nn
&+&\frac{1}{2}\big[ \eta_{\mu\nu}\big(p'\cdot p~ \eta_{\alpha\beta}-p_{\alpha}p'_{\beta}\big)+p'_{\{\mu}p_{\alpha}\eta_{\beta\nu\}}\\ 
&+&p'_{\beta}p_{\{\nu}\eta_{\mu\}\alpha}
-p'_{\{\mu}p_{\nu\}}\eta_{\alpha\beta}
-\big(p'\cdot p\big) \eta_{\{\mu\alpha} \eta_{\nu\}\beta}   \big]~.
\ea
\par
Once again, the expression above may be simplified using the standard techniques of 1-loop calculations. For brevity, again we present only the final result. The W-boson graviton coupling of  Fig. (\ref{gwc}) gives a contribution to the gauge current

\bea
\label{wgnc}
&&\bar{u}_{j} i \Delta J_{\mu \nu}^{(\text{\tiny{W}}^{\pm})}u_{i}\sim\sum_{l}\frac{i\kappa g^2}{16\pi^{2}} U_{lj}^* U_{li} \int_{R} dx~dy \\ \nn
&&\bigg[(3-2x-2y)\log\big[\Delta^{'}(q^{2},M_{W}^{2},m_{l}^{2})\big]\\ \nn
&-& \frac{xy q^{2}}{2\Delta^{'}(q^{2},M_{W}^{2},m_{l}^{2})}-\frac{(x+y)M^{2}_{W}}{[\Delta^{'}(q^{2},M_{W}^{2},m_{l}^{2})]^{2}}\bigg]\\ \nn
&&\bar{u}_{j}\gamma_{\{\mu}r_{\nu\}}\hat{P}_L u_{i}
\eea
where
\ba
\Delta^{'}(q^{2},M_{W}^{2},m_{l}^{2})=(x+y)M^{2}_{W}+(1-x-y)m^{2}_{l}-xy q^{2}~.
\ea

In the integral, $R$ denotes the region of the $X-Y$ plane bounded by the axes and the line $x+y=1$. Again, we note that the rank-2 tensor current comes out to have the required gauge invariant form. Note that the contribution from this diagram does have explicit $q^{2}$ dependences at this order. Hence we would expect it to contribute to the $q^{2}$ dependence of $E_{3}(q^{2})$ and the neutrino transition mass radius. The GIM mechanism~\cite{gim} again suppresses the contribution for $i\neq j$.

\begin{figure}
\begin{center}
\includegraphics[width=6cm,angle=0]{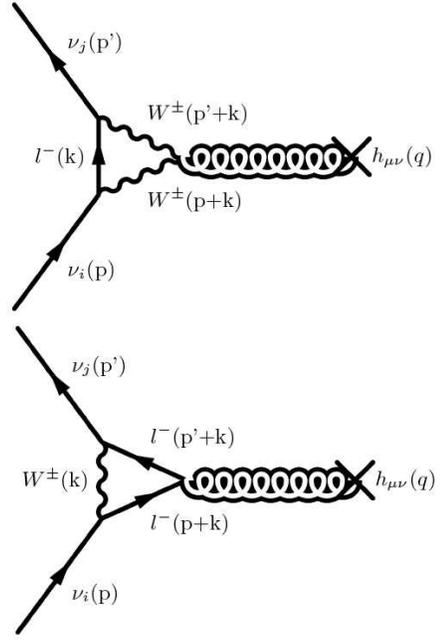}
\end{center}
\caption{Graviton-W boson and graviton-charged lepton coupled diagrams at 1-loop for the case $i \neq j$.}
\label{gwc}
\end{figure}

\par
The last Feynman diagram we consider is the one where the graviton is coupled to the charged lepton in the loop, as shown in Fig.\,(\ref{gwc}). A similar diagram also gives a contribution in the photon-neutrino coupling case.
For the diagram under consideration the tensor current is then given by
\ba
&&\bar{u}_{j}i\Delta J_{\mu \nu}^{(4)}u_{i}\sim\sum_{l=e,\mu,\tau} \frac{-\kappa g^2 U_{lj}^* U_{li}}{2} \int 
\frac{d^4k}{(2\pi)^4}\bar{u}_{j}(p')  \\ \nn
&&\frac{\left[ \gamma_{\alpha} P_L  (
\slashed{p}' + \slashed{k} + m_l) \Gamma^{''}_{\mu \nu} (\slashed{p} + \slashed{k} + m_l
) \gamma^{\alpha} \hat{P}_L\right]}{((p'+k)^2 -m_l^2) (k^2-M_W^2)((p+k)^2 -
m_l^2)}u_{i}(p) 
\ea
where
\ba
\Gamma^{''}_{\mu\nu}&=& \frac{1}{8}[\gamma_{\{\mu}(p+p')_{\nu\}}]
-\frac{1}{4}\eta_{\mu\nu}[\slashed{p}+\slashed{p}'-2m_{f}]~~.
\ea
\par
The above expression may be grouped in powers of $m_{l}$ and simplified. It is found that the only terms that survive are those proportional to $m^{2}_{l}$ and constant terms. Simplifying the expressions and performing the momentum integration finally gives
\bea
\label{lgcpl}
&&\bar{u}_{j} i \Delta J_{\mu \nu}^{(\text{l}^{-})}u_{i}\sim \sum_{l}\frac{i\kappa g^2}{64\pi^{2}} U_{lj}^* U_{li} \int_{R} dx dy~~~\\ \nn
&& (x+y-1)\bigg[-\frac{1}{2}\log\big[\Delta^{''}(q^{2},M_{W}^{2},m_{l}^{2})\big]-\\ \nn
&&\frac{m^{2}_{l}}{2\Delta^{''}(q^{2},M_{W}^{2},m_{l}^{2})}+\frac{(1-x-y+xy)q^{2}}{ \Delta^{''}(q^{2},M_{W}^{2},m_{l}^{2})}\bigg] \bar{u}_{j}\gamma_{\{\mu}r_{\nu\}}\hat{P}_L u_{i}
\eea
where
\ba
\Delta^{''}(q^{2},M_{W}^{2},m_{l}^{2})=(x+y)m^{2}_{l}+(1-x-y)M^{2}_{W}-xy q^{2}
\ea
Again in the Feynman integral, $R$ denotes the region of the $X-Y$ plane bounded by the axes and the line $x+y=1$.
 \par
Now we may put together the contributions from all the Feynman diagrams to get the net spin-2 tensor current. The combined 1-loop gauge current is  
\be
\label{nettc}
\Delta \hat{J}_{\mu \nu}^{\text{\,1-loop}}(q^{2})\sim\Delta \hat{J}_{\mu \nu}^{\text{\,4 pt.}}(q^{2})+\Delta \hat{J}_{\mu \nu}^{\,\text{\tiny{W}}^{\pm}}(q^{2})+\Delta\hat{ J}_{\mu \nu}^{\,\text{l}^{-}}(q^{2})
\eq
where $\Delta\hat{J}(q^{2})$ denotes the tensor current renormalized according to
\ba
\Delta \hat{J}_{\mu \nu}(q^{2})=\Delta{J}_{\mu \nu}(q^{2})-\Delta{J}_{\mu \nu}(0)
\ea
 in accordance with the conditions in Eq. (\ref{mreno}). Thus we observe that the contributions from the four-point graviton coupling in Fig.\,(\ref{4g1}) is renormalized away since it has no $q^{2}$ dependence at the order we are considering. The dominant contribution to $E_{3}(q^{2})$, $D_{3}(q^{2})$  and hence to the $i\rightarrow j$ transition mass radius, at this order of approximation, therefore comes from the Feynman diagrams in Fig. (\ref{gwc}). Also, as mentioned before, the off-diagonal contribution is suppressed as should be expected by the GIM mechanism~\cite{gim}. In the limit of degenerate charged lepton masses the 1-loop contributions would completely vanish and there would be no $i\rightarrow j$ transitions.
 \par
In Fig.(\ref{e3qsq}) we plot how, to leading order, the form factor $E_{3}(q^{2})\left[i\rightarrow j\right]$ changes with momentum transfer $-q^{2}$. It is seen that the $q^{2}$ dependence for moderate values of momentum transfer is very weak and almost insignificant. If there is any possibility of checking this effect in the gravitational scattering of neutrinos, one would have to consider some effect where the momentum transfer is very far from 0. For the scenarios we consider where $E_{\nu}\gg m_{\nu}$, note that 
\ba
q^{2}\approx~-~4E^{2}_{\nu}\sin^{2}\frac{\theta}{2}~~~.
\ea
Thus larger and larger values of momentum transfer correspond to larger and larger scattering angles $\theta$ for fixed values of $E_{\nu}$. 
\begin{figure}
\begin{center}
\includegraphics[width=9cm,angle=0]{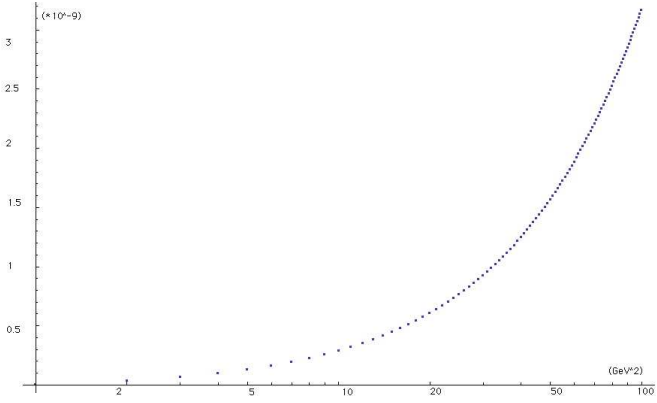}
\end{center}
\caption{$\vert E_{3}(q^{2}) \vert$ Vs. $-q^{2}$ plot for $i \neq j$ illustrating the approximate dependence to $q^{2}$ of the dominant transition form factor at lowest order. Note that the $q^{2}$ dependence is highly suppressed and we observe significant deviations from the vanishing $q\rightarrow~0$ value only at very large values of momentum transfer. }
\label{e3qsq}
\end{figure}
\par
Let us now estimate a neutrino transition mass radius based on Eq. (\ref{nettc}). Crudely speaking, the diagonal mass radius (in mass basis)  of a neutrino may be defined as the second moment of a hypothetical neutrino mass density
\be
\label{crd}
 \big\langle r^{2}_{\nu}\big\rangle_{m}\sim \int ~d^{3}r~\vert \vec{r}\vert^{2} ~\rho^{\text{\tiny{$\nu$}}}_{\text{\tiny{m}}}(\vec{r})
 \eq
Here as in section I, $\rho^{\text{\tiny{$\nu$}}}_{\text{\tiny{m}}}(\vec{r})$ is the fictitious neutrino mass density. If we assume that the hypothetical neutrino mass-density is spherically symmetric and that the momentum exchange $q$ is small then we may re-write the above expression as in Eq.(\ref{crdef}). This result may now be assumed to hold true even when Eq. (\ref{crd}) is not strictly correct and when $i\neq j$. This is again in parallel with the arguments leading to the definition of a neutrino charge radius. In our case then, a $\nu_{i}\rightarrow\nu_{j}$ transition mass radius may be defined as
 \ba
 \big\langle r^{2}_{\nu}\big\rangle_{m}\left[i\rightarrow j\right]\sim-6\frac{d}{dq^{2}}\left[E_{3}(q^{2})\right]\bigg\vert_{q^{2}\rightarrow 0}
 \ea
 \par
 $E_{2}(q^{2})$ is not generated in our leading order calculation and $E_{3}(q^{2})$ may be calculated from the expressions we obtain, after numerically evaluating the Feynman integrals. Using Eqs. (\ref{4ptc})-(\ref{nettc}) to calculate the above, say for $\nu_{2}\rightarrow \nu_{3}$, we find  the approximate value for the $\nu_{i}\rightarrow\nu_{j}$ transition mass radius
 \ba
  \big\langle r^{2}_{\nu}\big\rangle_{\text{\tiny{mass}}}\left[2\rightarrow 3\right]~\sim~1.5\times10^{-37}~\text{cm}^{2}
 \ea 
 \par
 A neutrino transition charge radius calculation with the same assumptions as the one we adopt (again for $2\rightarrow 3$)  gives
 \ba
\big\langle r^{2}_{\nu}\big\rangle_{\text{\tiny{charge}}}\left[2\rightarrow 3\right]~\sim~1.1\times10^{-33}~\text{cm}^{2}
 \ea
 \par
 Now, it is known that the neutrino charge radius is gauge dependent and hence unphysical, but there have been attempts to arrive at a gauge invariant definition for the neutrino charge radius. (See for example the relevant section in~\cite{pdg} and references therein.) Also, all the best-fit values quoted for the neutrino charge radius are in the ball-park of $10^{-32}-10^{-33}~\text{cm}^{2}$~\cite{pdg}. Our approximate calculation of a neutrino transition mass radius thus seems to indicate that 
 \ba
 \big\langle~ r^{2}_{\nu_{i}\rightarrow\nu_{j}}\big\rangle_{\text{\tiny{mass}}}~ <~\big\langle~ r^{2}_{\nu_{i}\rightarrow\nu_{j}}\big\rangle_{\text{\tiny{charge}}}
 \ea
 \par
 There is an intuitive way to understand the difference in order of magnitudes between the neutrino transition charge radius and neutrino transition mass radius. Let us consider the vacuum polarization of the neutrino. For the neutrino transition charge radius, the photon couples to the vacuum polarized $l^{-}$ and $W^{\pm}$ with equal ``weight", since they have the same magnitude of charge $\pm1$. Now, the definition of the Compton wavelength is
 \ba
 \lambda\simeq\frac{h}{mc}
 \ea
 where $h$ is the Planck's constant, $c$ is the speed of light and $m$ is the mass of the particle.
 Thus the charged lepton has a larger Compton wavelength compared to the $W^{\pm}$ boson and thereby it contributes more significantly to the neutrino charge radius. Taking a charge weighted average, we may associate heuristically 
 \be
 \big\langle r^{2}_{\nu}\big\rangle^{\lambda}_{\text{\tiny{charge}}}\sim\frac{\vert q_{l^{-}}\vert\lambda^{2}_{l^{-}}+\vert q_{W^{+}}\vert\lambda^{2}_{W^{+}}}{\vert q_{l^{-}}\vert+\vert q_{W^{+}}\vert}
 \label{rca}
 \eq
 where $q_{l^{-}}$ and $q_{W^{+}}$ denote the respective charges and $\lambda_{l^{-}}$ and $\lambda_{W^{+}}$ are the Compton wavelengths of  $l^{-}$ and $W^{+}$.
 \par
 Similarly, in the transition mass radius probed by the graviton, taking a weighted average with the respective masses, we may claim that
 \be
 \big\langle r^{2}_{\nu}\big\rangle^{\lambda}_{\text{\tiny{mass}}}\sim\frac{m_{l}\lambda^{2}_{l^{-}}+ M_{W}\lambda^{2}_{W^{+}}}{m_{l}+ M_{W}}
 \label{rma}
 \eq
 approximately.
  \par
Calculating the Compton wavelength for the charged leptons (appropriately weighted by the relevant PMNS matrix factors for the $2\rightarrow 3$ case) and the $W_{\pm}$, we get from Eqs.\, (\ref{rca}) and (\ref{rma}) 
 \ba
\frac{\big\langle r^{2}_{\nu}\big\rangle^{\lambda}_{\text{\tiny{mass}}}}{\big\langle r^{2}_{\nu}\big\rangle^{\lambda}_{\text{\tiny{charge}}}}~\approx~10^{-3}~~~.
 \ea
Thus, it is seen that the value calculated from heuristic arguments is in the ball park of the equivalent ratio calculated from 1-loop calculations,
\ba
\frac{\big\langle r^{2}_{\nu}\big\rangle^{\text{\tiny{1-loop}}}_{\text{\tiny{mass}}}}{\big\langle r^{2}_{\nu}\big\rangle^{\text{\tiny{1-loop}}}_{\text{\tiny{charge}}}}~\approx~10^{-4}~~~.
 \ea
 This probably gives a conceptual way of understanding the smallness of the neutrino mass radius compared to the neutrino charge radius.
\end{section}
\begin{section}{Summary}
\par
In this paper we studied the interaction of neutrinos with gravitational fields in the weak field regime. Among our aims were to understand the symmetries of the graviton-neutrino vertex in some detail and explore the gravitational neutrino transition form factors generated at 1-loop.
\par
The form of the graviton-neutrino vertex was motivated based on Lorentz and gauge invariance for both the mass diagonal and off-diagonal cases. The non-relativistic interpretations for the neutrino gravitational  form factors were derived. Also, renormalization criteria on the matrix elements in the mass basis were imposed so as to preserve the weak equivalence principle. We then used general arguments of spin-statistics, CP invariance and most importantly the symmetries of the angular momentum states to deduce theoretical differences between the Majorana and Dirac cases. Specific associations were made between the neutrino gravitational form factors and angular momentum states that are allowed. Differences in angular momentum states for the graviton case compared to the photon case were pointed out, for instance, the presence of states due to the $J=0$ part of the off-shell graviton. 
\par
It was then proved that, as in the electromagnetic case, the practical Majorana-Dirac confusion theorems are still valid in the gravitational case. This meant that in spite of the theoretical differences, due to the tiny neutrino masses and left-handed nature of the electroweak currents, the two cases (Majorana and Dirac) would be effectively indistinguishable for any space-time geometry satisfying the weak field condition. Finally we made some brief remarks about the results in sections II and III being valid for other Majorana particles that may be present in nature. 
\par
We then calculated the neutrino-graviton transition form factors generated at 1-loop to leading order and obtained an approximate understanding of their $q^{2}$ dependence. We pointed out that the gravity mediated $i\rightarrow j$ effect would be purely quantum mechanical and absent in the classical theory of linearized gravity. A neutrino transition mass radius in the presence of a gravitational field was then estimated from the approximate 1-loop expressions and compared to the neutrino transition charge radius. It was found from our calculation that the neutrino transition mass radius generally comes out to be smaller than the neutrino transition charge radius by couple of orders of magnitude. Finally, we tried to give a physical explanation for the smallness of the neutrino transition mass radius compared to the neutrino transition charge radius.
\par
There are many avenues left to be explored relating to the study of high energy neutrinos interacting with gravity. A more detailed calculation of the neutrino gravitational form factors with higher order terms included should clarify whether the behavior seen in our approximate computations for the neutrino transition form factors and neutrino transition mass radius is robust. Related to this is the question of gauge invariance of the gravitational form factors and whether they may even in principle be directly measured in some future neutrino experiment. This may be relevant in the context of exploring $CP$ phases in the lepton sector as we remarked in section III. Other speculations are whether we may use the differences in gravitational form factors of particles as cosmological probes, say for example to ascertain whether dark matter is bosonic or fermionic.
 \end{section}
\begin{acknowledgments}
 We are grateful to J. L. Rosner for many useful discussions and suggestions during this study, especially pertaining to angular momentum states and renormalization conditions, which were important to the present analysis. We thank B. Kayser and C. Wagner for interesting comments. We would also like to thank R. Akhoury, S. Farkas, S. Gralla, S. Green, O. Lunin and M. Seifert for discussions. A.M.T acknowledges support from the Subrahmanyan Chandrasekhar Fellowship during early stages of this work. This work was supported in part by the United States Department of Energy under Grant No. DE-FG02-90ER40560.
\end{acknowledgments}
\begin{appendix}
\section{Parity and charge conjugation properties for Dirac and Majorana two-particle final states.}
For completeness we briefly derive the well known P, C and CP properties of two particle fermion states in the Dirac and Majorana cases.
\subsection{Dirac two-particle final states}
For the bound state of two Dirac particles we may represent the final state to lowest order as
\begin{align*}
\vert \nu\bar{\nu};P, S \rangle_{D}=\sum_{s_{1},s_{2}}\int d^{3}k ~\Phi_{D}(\vec{k};s_{1},s_{2})~\hat{a}^{\dag}_{\vec{k},s_{1}}\hat{b} ^{\dag}_{-\vec{k},s_{2}}\vert 0\rangle+\ldots
\end{align*}
Let us first consider the effect of the parity operator $\mathcal{P}$ on the final state. Using the fact that~(see for example \cite{psch})
\begin{align*}
\mathcal{P}\hat{a}^{\dag}_{\vec{k},s_{1}}\mathcal{P}=\eta^{*}_{a} ~\hat{a}^{\dag}_{-\vec{k},s_{1}}\\
\mathcal{P}\hat{b}^{\dag}_{-\vec{k},s_{1}}\mathcal{P}=\eta^{*}_{b} ~\hat{b}^{\dag}_{\vec{k},s_{1}}
\end{align*}
and $\eta_{b}=-\eta_{a}^{*}$ we have after re-defining the variable of integration
\begin{align*}
\mathcal{P}\vert \nu \bar{\nu};P, S \rangle_{D}=\sum_{s_{1},s_{2}}\int d^{3}k^{'}~\Phi^{'}_{D}(\vec{k}^{'};s_{1},s_{2})~\hat{a}^{\dag}_{\vec{k}^{'},s_{1}}\hat{b} ^{\dag}_{-\vec{k}^{'},s_{2}}\vert 0\rangle \\
+\ldots
\end{align*}
In the above expression
\begin{align*}
 \Phi^{'}_{D}(\vec{k}^{'};s_{1},s_{2})=-\Phi_{D}(-\vec{k};s_{1},s_{2})
\end{align*}
Using the familiar properties of the spherical harmonics this gives the well known result
\begin{align*}
\zeta^{\text{\tiny{D}}}_{\text{\tiny{P}}} \left(^{2S+1}L_{J}\right) =(-1)^{L+1}
\end{align*}
Here $L$ and $S$ denote the angular momentum quantum number and total spin. $J$ is the total angular momentum defined in the usual way as $J=L+S$.
\par
Similarly, applying the charge conjugation operator $\mathcal{C}$ on the Dirac final state gives
\begin{align*}
\mathcal{C}\vert \nu\bar{\nu};P, S \rangle_{D}=\sum_{s_{1},s_{2}}\int d^{3}k^{''}~ \Phi^{''}_{D}(\vec{k}^{''};s_{1},s_{2})~\hat{a}^{\dag}_{\vec{k}^{''},s_{1}}\hat{b} ^{\dag}_{-\vec{k}^{''},s_{2}}\vert 0\rangle \\ 
+\ldots
\end{align*}
where
\begin{align*}
 \Phi^{''}_{D}(\vec{k}^{''};s_{1},s_{2})= -\Phi_{D}(-\vec{k};s_{2},s_{1})
\end{align*}
Using the properties of spherical harmonics and the spin states we get the familiar result
\begin{align*}
\zeta^{\text{\tiny{D}}}_{\text{\tiny{C}}} \left(^{2S+1}L_{J}\right) =(-1)^{L+S}
\end{align*}
\par
From the above two results it is easy to see that the CP parity for the Dirac two-particle final state is given by
\begin{align*}
\zeta^{\text{\tiny{D}}}_{\text{\tiny{CP}}} \left(^{2S+1}L_{J}\right) =(-1)^{S+1}
\end{align*}

\subsection{Majorana two-particle final states}
The bound state of the Majorana particles may be represented to lowest order again as
\begin{align*}
\vert \nu\nu;P, S \rangle_{M}=\sum_{s_{1},s_{2}}\int d^{3}k~ \Phi_{M}(\vec{k};s_{1},s_{2})~\hat{a}^{\dag}_{\vec{k},s_{1}}\hat{a} ^{\dag}_{-\vec{k},s_{2}}\vert 0\rangle\\
+\ldots
\end{align*}
where we have implicitly used the Majorana condition.
\par
Acting on the state by the parity operator $\mathcal{P}$ now gives
\begin{align*}
\mathcal{P}\vert \nu\nu;P, S \rangle_{D}=\sum_{s_{1},s_{2}}\int d^{3}k~ \Phi_{M}(\vec{k};s_{1},s_{2})~\eta_{a}^{*}\hat{a}^{\dag}_{-\vec{k},s_{1}}~\eta_{a}^{*}\hat{a} ^{\dag}_{\vec{k},s_{2}}\vert 0\rangle\\
+\ldots
\end{align*}
Unlike Dirac particles, Majorana particles have imaginary intrinsic parity. This is most easily seen by considering the expression $\mathcal{P}\Psi_{M}(\vec{x},t) \mathcal{P} ^{-1}$ and requiring it to be proportional to $\Psi_{M}(-\vec{x},t) $. This gives the condition
\begin{align*}
\eta_{a}^{*}=-\eta_{a}
\end{align*}
implying that $\eta_{a}=\pm i$. Using this condition, after a re-definition of the momentum integration, we get 
\begin{align*}
\mathcal{P}\vert \nu \nu;P, S \rangle_{M}=\sum_{s_{1},s_{2}}\int d^{3}k^{'}~ \Phi^{'}_{M}(\vec{k}^{'};s_{1},s_{2})~\hat{a}^{\dag}_{\vec{k}^{'},s_{1}}\hat{a} ^{\dag}_{-\vec{k}^{'},s_{2}}\vert 0\rangle \\
+\ldots
\end{align*}
where
\begin{align*}
 \Phi^{'}_{M}(\vec{k}^{'};s_{1},s_{2})=-\Phi_{M}(-\vec{k};s_{1},s_{2})
\end{align*}
which is similar to the Dirac case. This gives again
\begin{align*}
\zeta^{\text{\tiny{M}}}_{\text{\tiny{P}}} \left(^{2S+1}L_{J}\right) =(-1)^{L+1}
\end{align*}
\par
The action of the charge conjugation operator on the Majorana final state does not yield any new information since we have
\begin{align*}
\mathcal{C}\vert \nu \nu;P, S \rangle_{M}= \vert \nu \nu;P, S \rangle_{M}
\end{align*}
Thus
\begin{align*}
\zeta^{\text{\tiny{M}}}_{\text{\tiny{C}}} \left(^{2S+1}L_{J}\right) =+1
\end{align*}
\par
Putting the above results together we get finally for the CP parity of Majorana two-particle final states
\begin{align*}
\zeta^{\text{\tiny{M}}}_{\text{\tiny{CP}}} \left(^{2S+1}L_{J}\right) =(-1)^{L+1}
\end{align*}
\par
If the Majorana particles in the final state were two different eigenstates $a\neq b$ then in the above expressions $\eta^{*}_{a}\eta^{*}_{b}\neq-1$ and the CP parity would explicitly depend on the intrinsic parity of each particle. Thus in the off-diagonal case we would have
\begin{align*}
\zeta^{\text{\tiny{M}}}_{\text{\tiny{CP}}} \left(^{2S+1}L_{J}\right) =\eta^{*}_{a}\eta^{*}_{b}(-1)^{L}
\end{align*}
\par
From the above results we see that there are some key differences in the Majorana two-particle final state compared to the Dirac two-particle final state for charge conjugation, parity and CP.
\newpage
\section{Relevant Feynman rules derived for linearized gravity}
\par
\begin{figure} [h!]
\begin{center}
\includegraphics[width=5cm,angle=0]{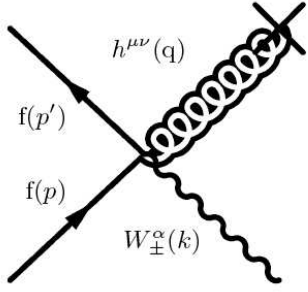}
\end{center}
\caption{For the four point graviton coupling the Feynman rule is $i \frac{\kappa g}{2\sqrt{2}}\big[\eta_{\mu\nu}\eta_{\alpha\beta}-\frac{1}{2}\eta_{\mu\beta}\eta_{\nu\alpha}-\frac{1}{2}\eta_{\mu\alpha}\eta_{\nu\beta}\big]\gamma^{\beta} \hat{P}_L$}
\end{figure}

\begin{figure} [th]
\begin{center}
\includegraphics[width=5cm,angle=0]{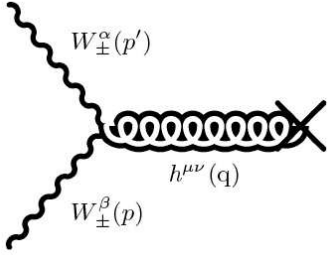}
\end{center}
\caption{ The Feynman rule is $ i\kappa \frac{M^{2}_{W}}{2}\big[\eta_{\mu\nu}\eta_{\alpha\beta}
-\eta_{\mu\alpha}\eta_{\nu\beta}-\eta_{\nu\alpha}\eta_{\mu\beta}\big]
-i\frac{\kappa}{2}\big[ \eta_{\mu\nu}\big(p'\cdot p~ \eta_{\alpha\beta}-p_{\alpha}p'_{\beta}\big)+p'_{\{\mu}p_{\alpha}\eta_{\beta\nu\}}
+p'_{\beta}p_{\{\nu}\eta_{\mu\}\alpha}-p'_{\{\mu}p_{\nu\}}\eta_{\alpha\beta}
-\big(p'\cdot p\big) \eta_{\{\mu\alpha} \eta_{\nu\}\beta}   \big]
$}
\end{figure}

\begin{figure}[h!]
\begin{center}
\includegraphics[width=5.5cm,angle=0]{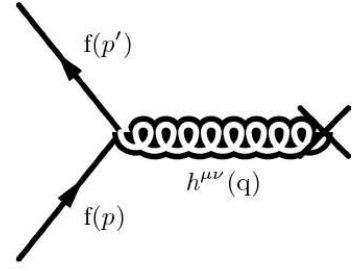}
\end{center}
\caption{Graviton-fermion coupling is $\frac{-i\kappa}{8}[\gamma_{\{\mu}(p+p')_{\nu\}}]
+\frac{i\kappa}{4}\eta_{\mu\nu}[\slashed{p}+\slashed{p}'-2m_{f}]$}
\end{figure}

\end{appendix}


\begin{thebibliography}{99}

\bibitem{astroneu}{K. Greisen, Phys. Rev. Lett. \textbf{16}, 748 (1966); G. T. Zatsepin and V. A. Kuzmin , JETP Lett. \textbf{4}, 78 (1966); V. S. Berezinsky and G. T. Zatsepin, Phys. Lett. \textbf{28B}, 423 (1969); J. Bahcall, \textit{Neutrino Astrophysics}, Cambridge Univ. press (1989); R.~J.~Protheroe, Nucl.\ Phys.\ Proc.\ Suppl.\  {\bf 77}, 465 (1999); S.~Pakvasa, Mod. Phys. Lett. A \textbf{23},1313 (2008); J.~F.~Beacom, arXiv:0706.1824 [astro-ph]; C.~Quigg, arXiv:0802.0013 [hep-ph]. }

\bibitem{anexp}{ A.~Silvestri {\it et al.}  [ANITA Collaboration], Mod.\ Phys.\ Lett.\  A {\bf 22}, 2237 (2007); S.~Razzaque, J.~A.~Adams, P.~Harris and D.~Besson, Astropart.\ Phys.\  {\bf 26}, 367 (2007) [arXiv:astro-ph/0605480]; H.~Landsman, L.~Ruckman and G.~S.~Varner  [IceCube Collaboration], {\it Prepared for 30th International Cosmic Ray Conference (ICRC 2007), Merida, Yucatan, Mexico, 3-11 Jul 2007}; S.~W.~Barwick, J.\ Phys.\ Conf.\ Ser.\  {\bf 60}, 276 (2007)[arXiv:astro-ph/0610631]; E.~Aslanides {\it et al.}  [ANTARES Collaboration],
  arXiv:astro-ph/9907432; S.~E.~Tzamarias  [NESTOR Collaboration], Nucl.\ Instrum.\ Meth.\  A {\bf 502}, 150 (2003); P.~Piattelli  [NEMO Collaboration], Nucl.\ Phys.\ Proc.\ Suppl.\  {\bf 143}, 359 (2005).
    }
\bibitem{brwh}{D. R. Brill and J. A. Wheeler, Rev. Mod. Phys., Vol. 29, 465 (1957).}

 \bibitem{voep}{M.~Gasperini, Phys.\ Rev.\  D {\bf 38}, 2635 (1988); A.~Halprin and C.~N.~Leung, Phys.\ Rev.\ Lett.\  {\bf 67}, 1833 (1991).}
  
  \bibitem{gphase}{D.~V.~Ahluwalia and C.~Burgard, Gen.\ Rel.\ Grav.\  {\bf 28}, 1161 (1996)
  [arXiv:gr-qc/9603008]; C.~Y.~Cardall and G.~M.~Fuller, Phys.\ Rev.\  D {\bf 55}, 7960 (1997)
  [arXiv:hep-ph/9610494]; N.~Fornengo, C.~Giunti, C.~W.~Kim and J.~Song, Nucl.\ Phys.\ Proc.\ Suppl.\  {\bf 70}, 264 (1999) [arXiv:hep-ph/9711494].}
  
   \bibitem{pbpjn}{J. F. Nieves and P. B. Pal, Phys. Rev. \textbf{D58}, 096005 (1998); J.~F.~Nieves and P.~B.~Pal, Mod.\ Phys.\ Lett.\  A {\bf 14}, 1199 (1999)
  [arXiv:gr-qc/9906006]; J.~F.~Nieves and P.~B.~Pal, Phys.\ Rev.\  D {\bf 63}, 076003 (2001) [arXiv:hep-ph/0006317].}
  
  \bibitem{maximdv}{A. Grigoriev, M. Dvornikov and A. Studenikin, Int. J. Mod. Phys. D \textbf{14}, p. 309 (2005), M. Dvornikov, Int. J. Mod. Phys. D \textbf{15}, p. 1017 (2006).}
  
    \bibitem{kay}{ B. Kayser, Phys. Rev. \textbf{D26}, 1662 (1982); J. F. Nieves, Phys. Rev. \textbf{D26}, 3152 (1982); B. Kayser and A. S. Goldhaber, Phys. Rev. \textbf{D28}, 2341 (1983). }
    
\bibitem{klak}{A.~Khare and J.~Oliensis,
   Phys.\ Rev.\  D {\bf 29}, 1542 (1984); K.~L.~Ng, Phys.\ Rev.\  D {\bf 47}, 5187 (1993)
  [arXiv:gr-qc/9305002].}
    
\bibitem{grb}{L. D. Landau and E. M. Lifshitz, \textit{Classical Theory of Fields}, Elsevier (2005); R. M. Wald, \textit{General Relativity}, Univ. of Chicago Press (1984); S. Weinberg, \textit{Gravitation and Cosmology}, Wiley (1972).}

\bibitem{Nieves:2007jz}
  J.~F.~Nieves and P.~B.~Pal,
  Phys.\ Rev.\  D {\bf 77}, 113001 (2008)
  [arXiv:0712.4345 [hep-ph]].
  
\bibitem{hp}{A. Pais and S.T. Epstein, Rev. Mod. Phys. \textbf{21}, 445 (1949); F. Rohrlich, Phys. Rev. \textbf{77}, 357 (1950); F. Villars, Phys. Rev. \textbf{79}, 122 (1950); S. Borowitz and W. Kohn, Phys. Rev. \textbf{86}, 985 (1952); H. Pagels, Phys. Rev. \textbf{144}, 1250 (1966).}

 \bibitem{ddsw}{R.~P.~Feynman, F.~B.~Morinigo, W.~G.~Wagner and B.~Hatfield, \textit{Feynman lectures on gravitation}, Reading, USA: Addison-Wesley (1995); H. van Dam and M. Veltman, Nucl. Phys. B22, 397 (1970); M.J.G Veltman, \textit{Quantum Theory of Gravitation}, North Holland Publ. (1970); D.~Dicus and S.~Willenbrock, Phys.\ Lett.\  B {\bf 609}, 372 (2005)
  [arXiv:hep-ph/0409316].}

\bibitem{dng}{D. Singh, N. Mobed and G. Papini, Phys. Rev. Lett. \textbf{97}, 041101 (2006); J. F. Nieves and P. B. Pal, in a criticism of the above paper,  Phys.\ Rev.\ Lett.\  {\bf 98}, 069001 (2007)
  [arXiv:gr-qc/0610098]; D. Singh, N. Mobed and G. Papini, in a reply to the above criticism, Phys.\ Rev.\ Lett.\  {\bf 98}, 069002 (2007),  [arXiv:gr-qc/0611016].}
  

\bibitem{ggibbons}
  G.~W.~Gibbons and M.~Rogatko,
  Phys.\ Rev.\  D {\bf 77}, 044034 (2008)
  [arXiv:0801.3130 [hep-th]].

  
\bibitem{psch}{S. Weinberg, \textit{The Quantum Theory of Fields}, Cambridge Univ. press (2002); M. E. Peskin and D. V. Schroeder, \textit{An Introduction to Quantum Field Theory}, Westview press (1995); T. Cheng and L. Li, \textit{Gauge Theory of Elementary Particle Physics}, Clarendon Press, Oxford (2005).}
  
 \bibitem{qftcs}{B. DeWitt, Phys. Rev. \textbf{D 162}, 1239 (1967); R. M. Wald, \textit{Quantum Field Theory in Curved Spacetime and Black Hole Thermodynamics}, Univ. of Chicago Press (1994); N. D. Birrell and P. C. W. Davies, \textit{Quantum Fields in Curved Spacetime}, Cambridge University Press (1994); S. Y. Choi, J. S. Shim and H. S. Song, Phys. Rev. \textbf{D51}, 2751 (1995).}
 
 \bibitem{mns}{B. Pontecorvo, Sov. Phys. JETP \textbf{6}, 429 (1957) ; Z. Maki, M. Nakagawa and S. Sakata, Prog. Theor. Phys. \textbf{28}, 870 (1962); B. W. Lee and R. E. Schrock, Phys. Rev. D \textbf{16}, 1444 (1977).}
 
\bibitem{gim}{S.~L.~Glashow, J.~Iliopoulos and L.~Maiani, Phys.\ Rev.\  D {\bf 2}, 1285 (1970).}
   
 \bibitem{pdg}{W.-M. Yao \textit{ et al.}, [Particle Data Group], J. Phys. G \textbf{33}, 1 (2006).}

\end {thebibliography}
\end{document}